\documentclass[12pt,3p,authoryear]{elsarticle}
\usepackage[T1]{fontenc}
\usepackage{amsmath,amssymb,amsfonts}
\usepackage{natbib}
\usepackage{graphicx}
\usepackage[hyphens]{url}
\usepackage[table]{xcolor}
\usepackage{epstopdf}
\usepackage{booktabs}
\usepackage{rotating}
\usepackage{times}
\usepackage{bm}
\usepackage{xspace}
\usepackage{color}
\usepackage{subdepth}
\usepackage{float}
\usepackage{setspace}
\usepackage{needspace}   
\usepackage[labelfont=bf,position=top]{caption}
\usepackage{geometry}
\usepackage[%
pdftex,%
pdfauthor={Me}%
pdfstartview={XYZ null null 1},%
pdffitwindow=true,%
pdfcreator={pdflatex}
letterpaper,%
colorlinks,%
breaklinks=true,%
backref=true,%
plainpages=false,%
bookmarks=false,%
pdfpagelabels,
pdfpagemode=None%
]{hyperref}

\newcommand{\grb}[1]{#1$^{*}$}
\newcommand{\grbb}[1]{#1$^{**}$}
\newcommand{\grbbb}[1]{#1$^{***}$}

\usepackage[font={small,singlespacing},labelfont={bf,small},skip=0.2cm]{caption}
\floatstyle{plaintop}
\restylefloat{table}
\restylefloat{figure}

\RequirePackage{ae,fancyvrb}
\newcommand{\pkg}[1]{\textbf{#1}}
\newcommand{\R}{\textsf{R}\xspace}

\journal{SSRN}

\usepackage{lineno}

\usepackage{adjustbox}
\usepackage{longtable}
\usepackage{tabularx,xltabular}

\newcommand{\ie}{\emph{i.e.}\xspace}
\newcommand{\eg}{\emph{e.g.}\xspace}

\newcommand{\insertfloat}[1]{%
\begin{center}
[Insert~#1 about here.]%
\end{center}%
}

\begin{document}
\begin{frontmatter}
\title{Thirty Years of Academic Finance\tnoteref{label1}}
\textsc{\tnotetext[label1]{We thank Kris Boudt, Jérôme Taillard, participants at the the 2022 SCSE annual meeting, the 2022 RFinance conference, and seminar participants at the Universities of Lausanne, Liège, and Sherbrooke for their comments. We are grateful to IVADO, the Natural Sciences and Engineering Research Council of Canada (grant RGPIN-2022-03767), and the Swiss National Science Foundation (grants \#179281 and \#191730) for their financial support.}
\author[hec]{David Ardia}
\ead{david.ardia@hec.ca}
\author[udes]{Keven Bluteau}
\ead{keven.bluteau@usherbrooke.ca}
\author[hec]{Mohammad-Abbas Meghani}
\ead{abbas.meghani@hec.ca}
\address[hec]{GERAD \& Department of Decision Sciences, HEC Montréal, Montréal, Canada}
\address[udes]{Department of Finance, Université de Sherbrooke, Canada}}

\begin{abstract}
We study how the financial literature has evolved in scale, research team composition, and article topicality across 32 finance-focused academic journals from 1992 to 2021. 
We document that the field has vastly expanded regarding outlets and published articles. Teams have become larger, and the proportion of women participating in research has increased significantly. Using the Structural Topic Model, we identify 45 topics discussed in the literature. We investigate the topic coverage of individual journals and can identify highly specialized and generalist outlets, but our analyses reveal that most journals have covered more topics over time, thus becoming more generalist. Finally, we find that articles with at least one woman author focus more on topics related to social and governance aspects of corporate finance. We also find that teams with at least one top-tier institution scholar tend to focus more on theoretical aspects of finance.
\end{abstract}
\end{frontmatter}

\doublespacing

\newpage
\section{Introduction} 

\noindent
Academic finance is an active research field that has evolved across several dimensions. Understanding this evolution can shed light on the domain's past, current, and near-future trends, and help finance scholars better position their research in the field. In this paper, we offer relevant takeaways for both newcomers to the field and more seasoned researchers, as well as journal editors, and practitioners. 

Specifically, we study how the financial literature has evolved in scale, research team composition, and article topicality. We look at these three dimensions to better understand changes in (i) the size of the field (number of outlets and articles), (ii) the research team's composition (size, gender, and affiliation), and (iii) the research subjects and their popularity, relations, and trends. To perform these analyses, we collect and analyze data from over 46,000 research articles published across 32 well-recognized finance-focused journals for the 1992--2021 period. 

In terms of size, we show that academic finance is a field that has vastly expanded. Over the last 30 years, the total number of articles published yearly has increased from 571 (by 19 outlets) to 3,617 (by 31 outlets). The average number of articles per outlet has increased from 30 to 117. The largest median growth across journals happened in the last two years of our sample, with an increase in publications of 18\% in 2020 and 27\% in 2021. We also find significant variability in publication growth across journals. For instance, comparing the average yearly publications in 1992--2005 and 2006--2021, four journals show a slight decrease (\emph{Journal of Finance}, \emph{Financial Analysts Journal}, \emph{Financial Review}, and \emph{Journal of Financial Research}), while three journals increased their publications by more than threefold (\emph{Finance Research Letters}, \emph{International Review of Financial Analysis}, and \emph{Journal of Corporate Finance}).

Regarding the research team dimension, we first show that the average number of authors per article has increased from 1.81 to 2.71. Second, we find that the percentage of articles with at least one female author has increased significantly from 15\%  to 45\%. Finally, scholars affiliated with a top-tier institution contribute to about 15\% to 23\% of all research papers published in our journal selection. For the three dimensions, we observe significant variations across journals. For instance, the average number of authors in articles published in the \emph{Pacific-Basin Finance Journal} is much higher than those published in the \emph{Journal of Money, Credit and Banking}. In addition, teams with women publish more in corporate finance outlets, such as the \emph{Journal of Corporate Finance}, than in the \emph{Journal of Portfolio Management}. Finally, the \emph{Journal of Finance}, \emph{Journal of Financial Economics}, and \emph{Review of Financial Studies} contain a much higher proportion of articles with at least one top-tier institution author.

Our primary analysis focuses on the subjects covered in the financial literature. Such a topic analysis is challenging as the relationships and evolution of financial themes covered in the literature are complex to grasp, topics are interrelated, their coverage differs among outlets, and their popularity (or prevalence) is not static. Moreover, article characteristics such as the team's composition can be related to the research topic. We address that challenge using the Structural Topic Model (STM) of \citet{RobertsEtAl2013}. The STM belongs to the class of topic models that seek to discover latent thematic clusters within a collection of texts. Besides identifying such groups, topic models estimate the prevalence of the found topics. STM presents several advantages compared to the seminal and widely used Latent Dirichlet Allocation (LDA) approach by \citet{BleiEtAl2003}. First, it accounts for the potential correlation between the topics' prevalence, likely in the finance domain, as subjects are often related. Second, it allows researchers to estimate the relation between topics and document-level information, such as year of publication or outlet, as in our application.
Note that while STM is commonly used in political science \citep[\eg,][]{MishlerEtAl2015,SachdevaEtAl2017,CurryFix2019,BohrDunlap2018}, examples from finance are much rarer;  see \citet{ArdiaEtAl202x} for a recent application in the financial context. We apply the STM to research articles' titles and abstracts. We use covariates in the model to account for the fact that topics' prevalence can depend on time, outlet, and research team composition. Our estimations lead to the identification of 45 topics in our corpus, from which we can draw several observations.

First, using network analysis, we find that several topics are strongly linked and can be grouped into clusters and show how these clusters are related together. In particular, we can identify the clusters ``corporate finance'', ``banking and insurance'', ``financial risk'', and ``financial market and asset pricing''.  Topic \textsf{Debt Financing} links ``corporate finance'' and ``banking and insurance'', \textsf{Risk Management}, \textsf{Monetary Policy}, and \textsf{Bond Market} bridge ``banking and insurance'' to ``financial risk'', \textsf{Portfolio Stategy} links ``financial risk'' to ``financial market and asset pricing'', and  \textsf{Analyst} and \textsf{Firm Valuation} link ``financial market and asset pricing'' and ``corporate finance''. 

Second, we look at the highest- and lowest-prevalent topics over time. We find that financial market-related topics such as \textsf{Market Anomaly} and \textsf{Trading/Liquidity} were predominant in the 1992--2008 period. Following the 2008 financial crisis and until 2016, \textsf{Monetary Policy} and \textsf{Bank Liquidity} were the most prevalent topics. In recent years,  \textsf{Corporate Social Responsibility} has become the leading subject. Furthermore, a trend analysis reveals that low-prevalent topics in the 1990s have experienced an increase in prevalence (\eg, \textsf{Household Credit/Economic Crash} and \textsf{Political Relationship/Corruption}). In contrast, high-prevalent topics have experienced a decrease (\eg, \textsf{Option Pricing} and \textsf{Volatility}). These results suggest that the financial literature has become more diversified over time.

Third, we investigate the topic coverage of individual journals and can identify highly specialized and more generalist outlets. For instance, more than 34\% of the papers published in the \emph{Journal of Derivatives} are related to \textsf{Option Pricing}, while the most prevalent topic in the \emph{European Journal of Finance} accounts for only 3.01\% of the topical distribution. Our analyses reveal that most journals have covered more topics over time, thus becoming more generalist. An exception is the \emph{Financial Analysts Journal}, which is more specialized nowadays. Links are also observed between journals according to the topic they cover. For instance, the commonly recognized top-five finance journals are close regarding topic coverage.

Finally, we analyze how topic prevalences are related to the research team characteristics. In particular, we find that articles with at least one woman author focus more on topics related to social and governance aspects of ``corporate finance.'' On the contrary, the prevalence is less important for topics oriented toward technical and modeling aspects in finance. We also find that teams with at least one top-tier institution scholar tend to focus more on theoretical aspects of finance.

Our paper contributes to the emerging literature on scientometrics in economics and finance that uses natural language processing tools as their basis for analysis. \citet{AmbrosinoEtAl2018} investigate topics in economic journals using multiple LDA models spanning several decades.  \citet{CorbetEtAl2019} also uses LDA to study the literature on the financial economics of precious metals. \citet{AlexakisEtAl2021} analyze the major topics published in \emph{Computational Economics} and find the 18 relevant topics covered by the journal over the 1993--2019 period. In \citet{BerningerEtAl2021}, as part of their study, they analyze 16 finance journals using LDA and find that the prevalence of topics remains relatively constant over time. Our use of STM is more efficient as it is estimated over the whole corpus and accounts for changes in prevalence conditional on the date by construction.\footnote{Also, applying LDA on a rolling window can lead to different topic-word distributions resulting in the topics themselves being different.} Accounting for temporal effects using STM, our results contrast with \citet{BerningerEtAl2021} as we observe that most topics have gained or lost prevalence over time. \citet{dai2021dissemination}, using articles posted on the Financial Economics Network (on SSRN), analyze how the characteristics of research articles (including team composition) are related to research outcomes (\eg, conference acceptance, top-three publication). \citet{grossmann2022analysis} analyze the relationship between journal accessibility, quality, and regional representation (proxying for inclusivity). \citet{BakerEtAl2021} perform a scientometric analysis of the \emph{Journal of Corporate Finance} and identify topics based on author keywords. Finally, \citet{conde2021gender} identify topics in the top 5 economic journals and analyze the relationship between the author gender and research topic using the STM model. 

The rest of this paper is structured as follows. In Section~\ref{sec:data}, we describe our data collection process, variables, and text-cleaning procedures. In Section~\ref{sec:stm}, we introduce the STM. The results are discussed in Section~\ref{sec:results}. Section~\ref{sec:conclusion} concludes.

\section{Data Collection, Text Processing, and Covariates}
\label{sec:data}

\noindent
This section presents the data collection process, the text processing steps, and the covariates used in our analyses.

\subsection{Data Collection}

\noindent
We retrieve data for 32 financial journals from the Web of Science Core Collection (WOC-CC) and Scopus.\footnote{See \url{https://clarivate.com/webofsciencegroup/solutions/web-of-science-core-collection} and \url{https://www.scopus.com}.} The outlets' selection is based on the list compiled in \citet{BajoEtAl2020}. We collect articles' titles, abstracts, and metadata for each journal from January 1992 to December 2021. We then merge the two databases using the digital object identifier (DOI) to obtain a complete list of articles. Across databases, we keep only articles classified as "Article." This step removes documents such as reprints, book reviews, letters, and corrections. These steps yield a corpus of 46,144 articles across all 32 journals. Although our corpus consists only of the titles and abstracts, we refer to each data point as an ``article'' in what follows for convenience.\footnote{Our dataset contains missing data for a few journal-year pairs: \emph{Review of Finance} 2003, \emph{European Journal of Finance} 1997--1999, \emph{Journal of Money, Credit and Banking} 1993--1996, \emph{European Journal of Finance} 1997, and \emph{Journal of Portfolio Management} 2018. In these cases, the information was unavailable on the databases, or the abstract was missing. Extrapolating from the number of articles available for these outlets in the surrounding years, we find that the missing data represent approximately 0.3\% (130 articles) of the sample and should therefore not influence the conclusion of our analyses.}

In Table \ref{tab:journal}, we report the name of the journals, the number of articles collected, and the year of the first and last article available in our sample. With 4,884 articles, the \emph{Journal of Banking \& Finance} is the most represented outlet in our sample. It is followed by the \emph{Journal of Financial Economics} (2,750 articles), the \emph{Journal of Finance} (2,330 articles), and the \emph{Journal of International Money and Finance} (2,296 articles). The least represented outlet is the \emph{Journal of Business} (461 articles), followed by the \emph{Journal of Derivatives} (581 articles), and the \emph{Journal of Financial Markets} (588 articles).\footnote{We note that the \emph{Journal of Business} ceased publication at the end of 2006.} The most active outlets in terms of yearly publications are the \emph{Journal of Banking \& Finance} (163 articles/year), \emph{Finance Research Letters} (94 articles/year), and the \emph{Journal of Financial Economics} (92 articles/year). The least active outlets are the \emph{Journal of Derivatives} (20 articles/year), the \emph{Journal of Financial Intermediation} (21 articles/year), and the \emph{Journal of Financial Services Research} (21 articles/year).

\insertfloat{Table~\ref{tab:journal}}

\subsection{Text Processing}

\noindent
We build our text corpus by concatenating each article's title and abstract. These components represent summaries of the article's content and provide our topic model with enough data to obtain reliable results.\footnote{Most academic journals require the abstract to briefly state the purpose of the research, the principal results, and major conclusions. As such, the abstract summarizes the most relevant part of an article. Note also that we tried to include author-defined keywords in the modeling, but this did not improve the topics' interpretability.} To normalize each article, we process the texts as follows:

\begin{enumerate}
\item We remove non-informative components: numbers and punctuations, one-letter words, and extra spaces. 

\item To reduce the complexity of the number of individual features (\ie, the dictionary of tokens), we lemmatize each word into its root form. Stemming each word is an alternative, but \citet{SchofieldMimno2016} show it is not helpful in the context of topic modeling. 

\item We detect and combine collocation in our corpus to better identify topics from our articles. For instance, the compound words ``stock market'' or ``exchange rate'' provide a topic model with more information about the content of an article than if these words were not considered a combination. We use two methods to detect collocations in our documents: (i) RAKE \citep{RoseEtAl2010} and (ii) the process described in  \citet{HansenEtAl2018}. RAKE uses a machine-learning model to score candidate collocations and then combine them, giving a candidate list. The method by \citet{HansenEtAl2018}
looks at part of speech patterns within the text.\footnote{Following \citet{JustesonKatz1995} these patterns are: adjective-noun, noun-noun, adjective-adjective-noun, adjective-noun-noun, noun-adjective-noun, noun-noun-noun, and noun-preposition-noun. We find the part of speech of each word using the UDPipe methodology implemented in the \R package \pkg{udpipe} \citep{WijffelsEtAl2021}.} We tabulate the number of times each candidate collocation appears within the overall corpus. Then, for both methods, we keep the collocations that occur at least 100 (50) times for the sequence of two words (three words). Finally, we concatenate the individual words of the collocation in each text of our corpus (\eg, ``stock market'' becomes ``stock\_market''). 

\item We follow \citet{MartinJohnson2015} who find that removing all words except nouns improves downstream tasks such as topic modeling. Moreover,  abbreviations (\eg, ``CSR'' for ``Corporate Social Responsibility'') are classified as proper nouns and are highly indicative of the topic of research in an article. As such, we keep only the words in our article that are classified as either a noun, a proper noun, or a part of a collocation (as identified in the previous step).

\item Removing words that are too rare or too common in the text corpus effectively narrows down pertinent words inside the corpus. \citet[p.~273]{GrimmerStewart2013} suggest removing words that appear in less than 1\% or more than 99\% of the texts. As our textual data can be technical, we set the lower threshold to 0.1\%.

\item Finally, we transform the processed texts into a document-term matrix (DTM). A DTM is a matrix where each column is a token (a noun, pronoun, or a collocation in our case), and each row is a document. Each element corresponds to the number of times a token is observed for a given text. This matrix serves as the input for the STM. The matrix contains 46,144 rows (corpus size) and 3,866 columns (vocabulary size). In the online Appendix, Section~\ref{sec:vocabulary}, we report the complete list of tokens in our vocabulary, the number of times they appear, and the percentage of documents in which they are present.
\end{enumerate}

\subsection{Covariates}\label{sec:covariates}

\noindent
In addition to the title and the abstract of an article, we will use several additional document-level metadata in our modeling. 

\paragraph{Year}

The year of publication is a critical covariate for the trend analysis of topics. For instance, new financial innovations or changing socio-economic situations may lead to emergent topic research. 

\paragraph{Journal}

Academic finance journals typically target articles within a specific scope. The editorial team ensures that the article fits the scope
before starting the refereeing process. As such, the academic outlet in which an article is published is likely one of the most direct indicators of research topicality.

\paragraph{Number of authors}

Topicality may be related to the number of authors contributing to a research article. For example, a larger team of authors might be more suited to more complex and cross-disciplinary research topics. 

\paragraph{Gender}

The gender covariate indicates if at least one of the authors of an article is a woman. Several studies have shown that women's research interests are different than men's; see, for instance, \citet{dolado2012men}, \citet{chari2017gender}, and \citet{conde2021gender} in the discipline of economics. To determine the gender of an author, we use the author's full name in ``Gender API\footnote{See \url{https://gender-api.com/.}},'' which has been shown in \citet{santamaria2018comparison} to be the best performer among gender identification methods.

\paragraph{Top-Tier Finance Institution}

The top-tier institution covariate indicates if at least one of the authors of an article is affiliated with a top-tier institution in finance. We obtain a list of 25 top-ranked institutions from the QS University Ranking in the financial area.\footnote{As of June 2021, these institutions are (in decreasing ranking order): Harvard University, Massachusetts Institute of Technology, Stanford University, London School of Economics, University of Oxford, University of Cambridge, University of Pennsylvania, University of California, University of Chicago, New York University, London Business School, Columbia University, National University of Singapore, Bocconi University, Yale University, University of California Los Angeles, The University of Manchester, The University of Melbourne, The University of New South Wales, HEC Paris School of Management, The University of Sydney, The Hong Kong University of Science and Technology, Tsinghua University, Peking University, Nanyang Technological University. See \url{https://www.topuniversities.com/university-rankings/university-subject-rankings/2020/accounting-finance}. Note that the earlier version of this list available on the website is from 2019. We acknowledge that the ranking may change over time, but we use the 2021 list as a proxy for top institutions, given the persistence of such status over time.}

\section{Structural Topic Model}
\label{sec:stm}

\noindent
The STM belongs to the class of topic models that seek to discover latent thematic clusters within a collection of texts. Besides identifying such groups, topic models estimate the prevalence of the found topics. The STM is an extension of the latent Dirichlet allocation (LDA) model of \citet{BleiEtAl2003} and the correlated topic model (CTM) of \citet{BleiLafferty2006}. The STM incorporates in both models conditioning document-level covariates (\eg, time of publication, outlet) that can affect the topic prevalence.\footnote{Within the STM, document-level covariates can also be used to fine-tune topic-word distributions \citep{RobertsEtAl2016}. We do not consider this here, and rely on a homogeneous distribution over words representing each topic; see point 2. below.} 

\subsection{Generative Process Representation}

\noindent
Following the presentation in \citet{RobertsEtAl2019}, the generation of each document (indexed by $d \in \{1,\ldots,D\}$) with a vocabulary of size $V$ and $K$ topics can be summarized as follows:

\begin{enumerate}
\item Draw the document-level prevalence of each topic, a vector $\theta_d$ of size $K$, from a logistic-normal generalized linear model based on a vector of document covariates $x_d$:
\begin{equation} 
\theta_d | X_d \gamma,\Sigma  \, \sim \, \textit{LogisticNormal}(x_d\gamma,\Sigma)\,,
\end{equation}
where $x_d$ is a $1 \times p$ vector of covariates, $\gamma$ is a $p \times (K - 1)$ matrix of coefficients and $\Sigma$ is a $(K - 1) \times (K - 1)$ covariance matrix.

\item Form the homogeneous distribution over words representing each topic $k$, a vector $\beta_{k}$ of size $V$, using the baseline word distribution $m$ and  the topic-specific deviation $\kappa_k$:
\begin{equation} 
\beta_{k} \propto \exp(m +\kappa_k)\,.
\end{equation}
\item For each word $n \in \{1, . . . , N_d\}$ in the document $d$:
\begin{itemize}
\item Draw word’s topic assignment based on the document-specific distribution over topics:
\begin{equation} 
z_{d,n} | \theta_d \, \sim \, \textit{Multinomial}(\theta_d)\,.
\end{equation}
\item Conditional on the topic chosen, draw an observed word from that topic:
\begin{equation} 
w_{d,n} |z_{d,n}, \beta_{k=z_{d,n}} \, \sim \, \textit{Multinomial}( \beta_{k=z_{d,n}})\,.
\end{equation}
\end{itemize}
\end{enumerate}

\subsection{Document-Level Covariates}

\noindent
As emphasized, one major advantage of the STM over other topic models is that it can directly incorporate into its modeling framework prevalence-conditioning variables to each document $d$ via the vector $x_d$. In particular, for the covariates in $x_d$, we use:

\begin{itemize}
\item B-spline basis functions to transform the publication year into ten continuous features representing time, as in \citet{RobertsEtAl2016}.
\item A set of journal dummy variables.
\item The number of authors.
\item A dummy variable indicating whether one of the authors is a woman.\footnote{We could not identify the gender for about 1\% of the articles in our sample. To mitigate the bias in our model due to these missing data, we included a dummy variable indicating whether the gender could be identified with a confidence level.}
\item A dummy variable indicating whether one of the authors is affiliated to a top-tier institution in finance.
\end{itemize}

\subsection{Estimation and Covariates/Topic Relationships Testing}
\label{sec:testing}

\noindent
The STM is estimated with a partially-collapsed variational expectation-maximization algorithm; we refer to \citet{WangBlei2003} and \citet{RobertsEtAl2016} for further details.\footnote{We rely on the  \R package \pkg{stm} \citep{RobertsEtAl2019} for the STM estimation. We use the regularizing prior distributions for $\gamma$, $\kappa$, and $\Sigma$, which helps to enhance interpretation and prevent overfitting as advocated in \citet{RobertsEtAl2016}.} The estimation yields an approximate posterior distribution for the parameters of interest. First, it provides a topic prevalence matrix, $\Theta$, of size $D \times K$, where $D$ is the total number of articles (\ie, 46,144) and $K$ is the number of topics. Each row of that matrix is represented by $\theta_d$ with elements  $\theta_{d,k} \geq 0$ and $\theta_{d,K} = 1 - \sum_{k=1}^{K-1} \theta_{d,k}$. Second, it provides a matrix of word distributions for each topic, $B$, of size $K \times V$, where $V$ is the vocabulary size (\ie, 3,866). Each row of that matrix is represented by $\beta_k$ with elements  $\beta_{v,k} \geq 0$ and $\beta_{V,k} = 1 - \sum_{v=1}^{V-1} \beta_{v,k}$. Traditionally, a point estimate is used to summarize these two quantities, such as the maximum a posteriori (MAP). We will use the MAP in the descriptive analyses of the topics.

In situations where we want to test a relation between $\theta_{d,k}$ and document-level covariates in $x_d$, we must account for the fact that topics are latent and thus use the information contained in the posterior 
distribution  \citep{RobertsEtAl2016,SchulzeEtAl2021}. In the STM, \citet{RobertsEtAl2016} employ a variant of the method of composition proposed by \citet{TreierJackman2008}. They rely on Monte Carlo simulations in which they repeatedly sample from the posterior and estimate a linear model between the generated $\theta_{d,k}$ and $x_d$. To test specific relations between (functions of) $\theta_{d,k}$ and various specifications in $x_{d}$, we will adapt their approach as in \citet{SchulzeEtAl2021} by replacing the linear regression with a beta regression that assumes the dependent variable $\theta_{d,k}$ in the $(0,1)$ interval \citep{FerrariCribariNeto2004}.\footnote{We use the function \texttt{estimateEffect} from the \R package \pkg{stm} \citep{RobertsEtAl2019} in conjunction with the function \texttt{betareg} from the \R package \pkg{betareg} \citep{CribariNetoZeileis2010}. The standard errors and p-values are obtained from the method of composition that incorporates the STM estimation uncertainty.} 

\section{Results}
\label{sec:results}

\noindent
We present the empirical results in what follows. In Section~\ref{sec:size}, we detail the evolution of the size of the field in terms of publications and outlets. In Section~\ref{sec:authorship}, we look at the change in the research teams. In Section~\ref{sec:topicality}, we perform an in-depth analysis of the topicality of financial research based on an estimated STM. 

\subsection{Size of Research Field}\label{sec:size}

\begin{figure}[H]
\centering
\includegraphics[width=1\textwidth]{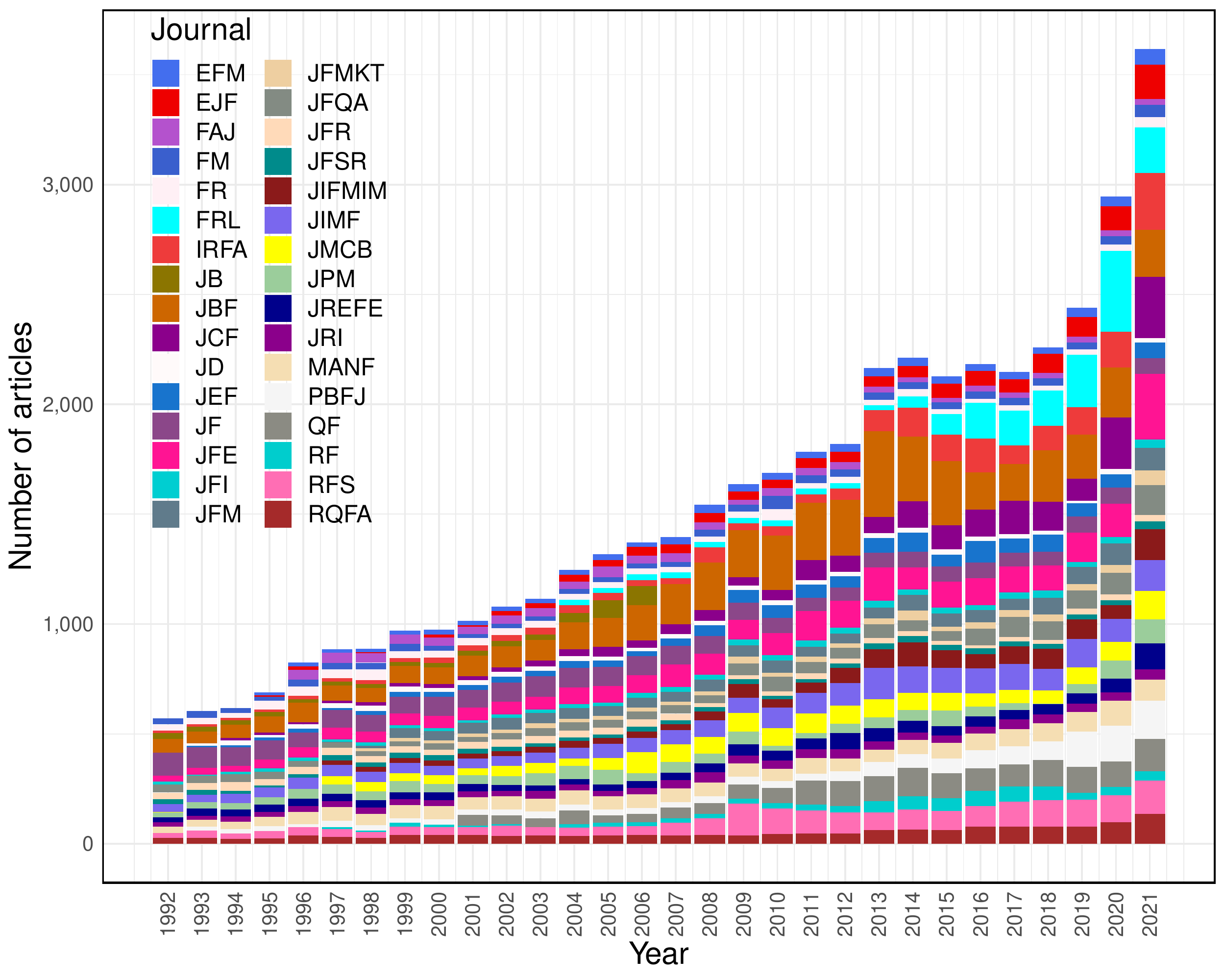}
\caption{\textbf{Number of Articles Published Over Time}\\
This figure displays the number of articles published by the 32 journals in our corpus during the 1992--2021 period.} 
\label{fig:journal}
\end{figure}

\noindent
In Figure~\ref{fig:journal}, we display the yearly number of publications per journal in our sample. We note the significant increase in publications over time, from 571 articles in 1992 to 3,617  in 2021. This increase is explained by: (i) an increased number of outlets in our sample, from 19 in 1992 to 31 in 2021, and (ii) an increased number of publications per outlet, from a yearly average of 30 publications in 1992 to 117 in 2021.\footnote{All journals but the \emph{Journal of Business} were in activity in December 2021.} The most significant median growth across journals happened in the last two years of our sample (an increase in publications of 18 \% in 2020 and 27\% in 2021). On aggregate, the field seems to have vastly expanded in terms of research output over the years. 

Focusing on individual journals, the yearly growth rate of publications averages 5.56\% (median at 4.36\%) and is positive for all but four journals (\ie, \emph{Journal of Finance}, \emph{Financial Analysts Journal}, \emph{Financial Review}, \emph{Journal of Financial Research}). The most significant increase is observed for the \emph{Journal of Financial Markets} with a yearly growth rate of 20.95\%, followed by \emph{Finance Research Letters} (16.94\%), and the \emph{Review of Finance} (14.45\%). The negative growth rates range from -2.05\% for the \emph{Journal of Finance} to -0.24\% for the \emph{Journal of Financial Research}. We note that the journals with large (low) growth tend to be relatively young (old) journals.

\insertfloat{Figure~\ref{fig:journal}}

\subsection{Characteristics of Research Teams}\label{sec:authorship}

\begin{figure}[H]
\centering
\includegraphics[width=1\textwidth]{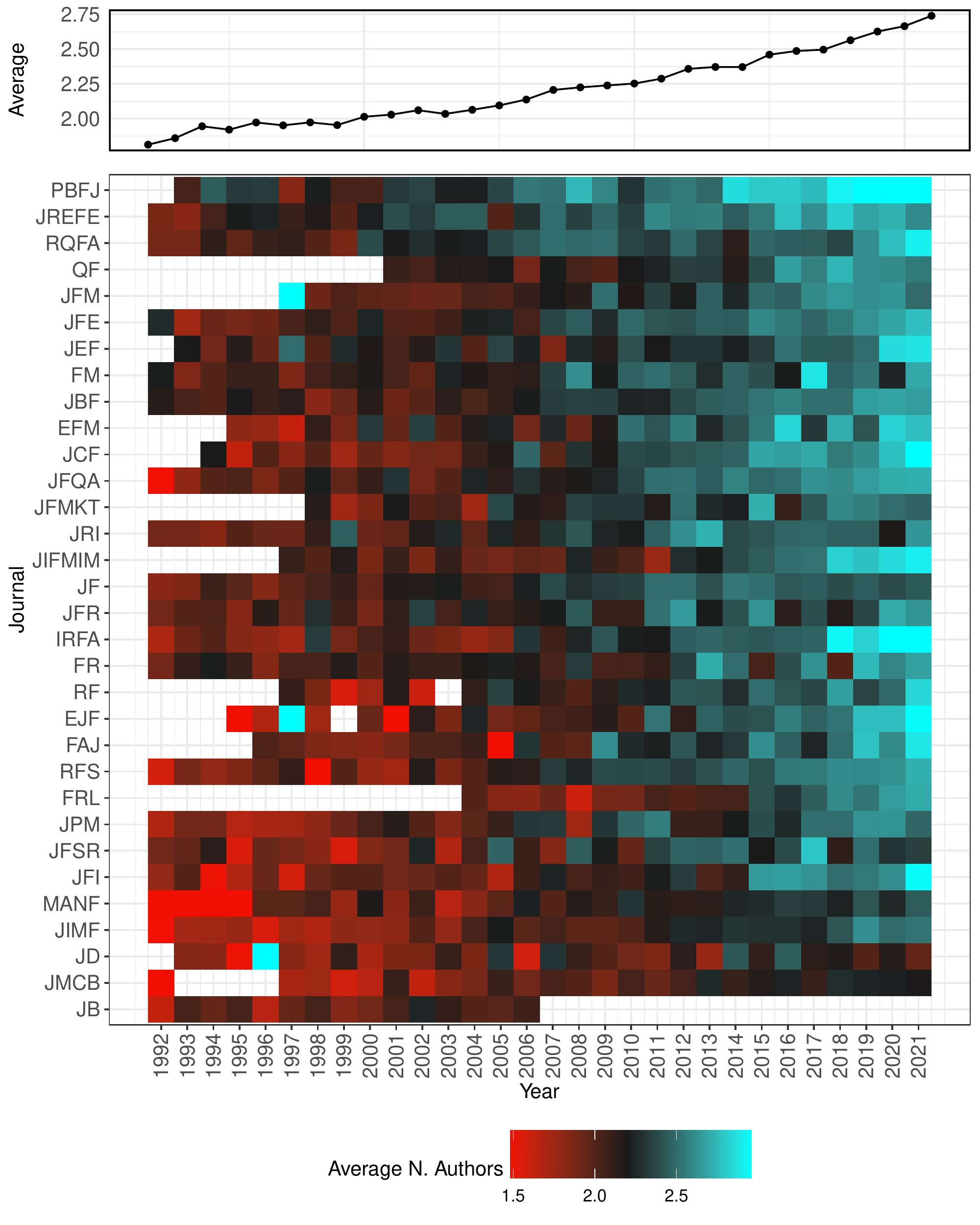}
\caption{\textbf{Average Number of Authors per Article Over Time}\\
This figure displays the average number of authors per article over time. The top panel shows the average over all outlets, while the 
bottom panel reports results per journal. Journals are sorted from higher to lower average number of authors over the 1992--2021 period.}
\label{fig:nauthors}
\end{figure}

\noindent
In Figure~\ref{fig:nauthors}, we display the evolution over the 1992--2021 period of the average number of authors per article. This number has increased from 1.81 to 2.71. The increasing pattern is present for almost all journals except the \emph{Journal of Derivatives} and the \emph{Journal of Money, Credit and Banking}. The journals with the largest average of authors per article are the \emph{Pacific-Basin Finance Journal} (2.48 authors) and the \emph{Journal of Real Estate Finance and Economics} (2.37 authors). We note that \emph{Pacific-Basin Finance Journal} focuses exclusively on Asia-Pacific countries, which may require more co-authors, given data acquisition and country-specific complexities. Similarly, the \emph{Journal of Real Estate Finance and Economics} focuses on real estate from finance and economic perspectives, which may require different expertise leading to a larger research team. 

\insertfloat{Figure~\ref{fig:nauthors}}

\begin{figure}[H]
\centering
\includegraphics[width=1\textwidth]{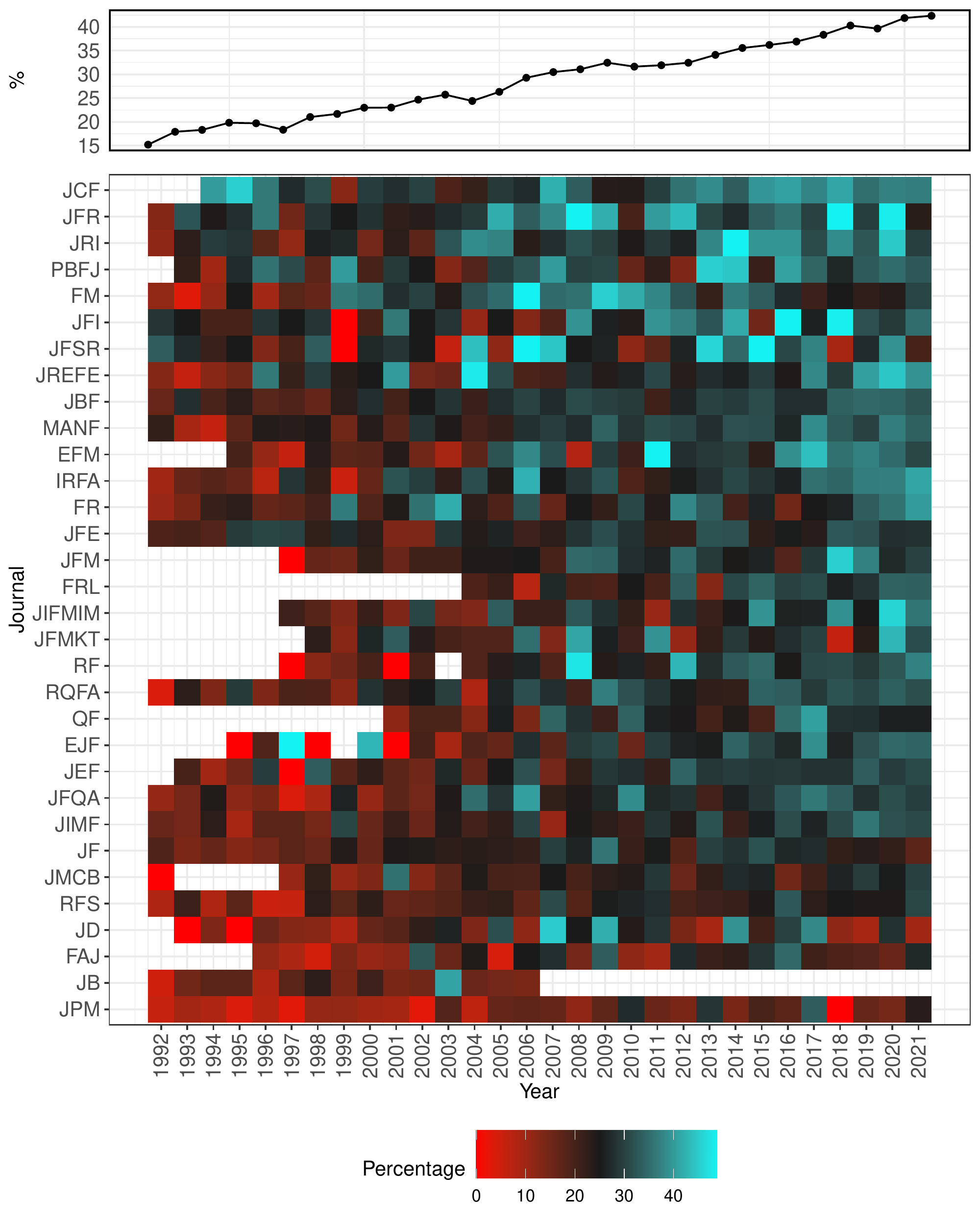}
\caption{\textbf{Proportion of Articles With at Least One Female Author Over Time}\\
This figure displays the proportion of articles with at least one female author. The top panel shows the proportion over all outlets, while the 
bottom panel reports results per journal. Journals are sorted from higher to lower proportions computed over the 1992--2021 period.}
\label{fig:women}
\end{figure}

In Figure \ref{fig:women}, we show the evolution of the proportion of articles with at least one female author. This proportion has increased steadily from around 15\% in 1992 to 45\% in 2021. Also, we note variations across journals: The average proportion is 33\% for the \emph{Journal of Corporate Finance} while it is 15\% for the \emph{Journal of Portfolio Management}. 

\insertfloat{Figure~\ref{fig:women}}

\begin{figure}[H]
\centering
\includegraphics[width=1\textwidth]{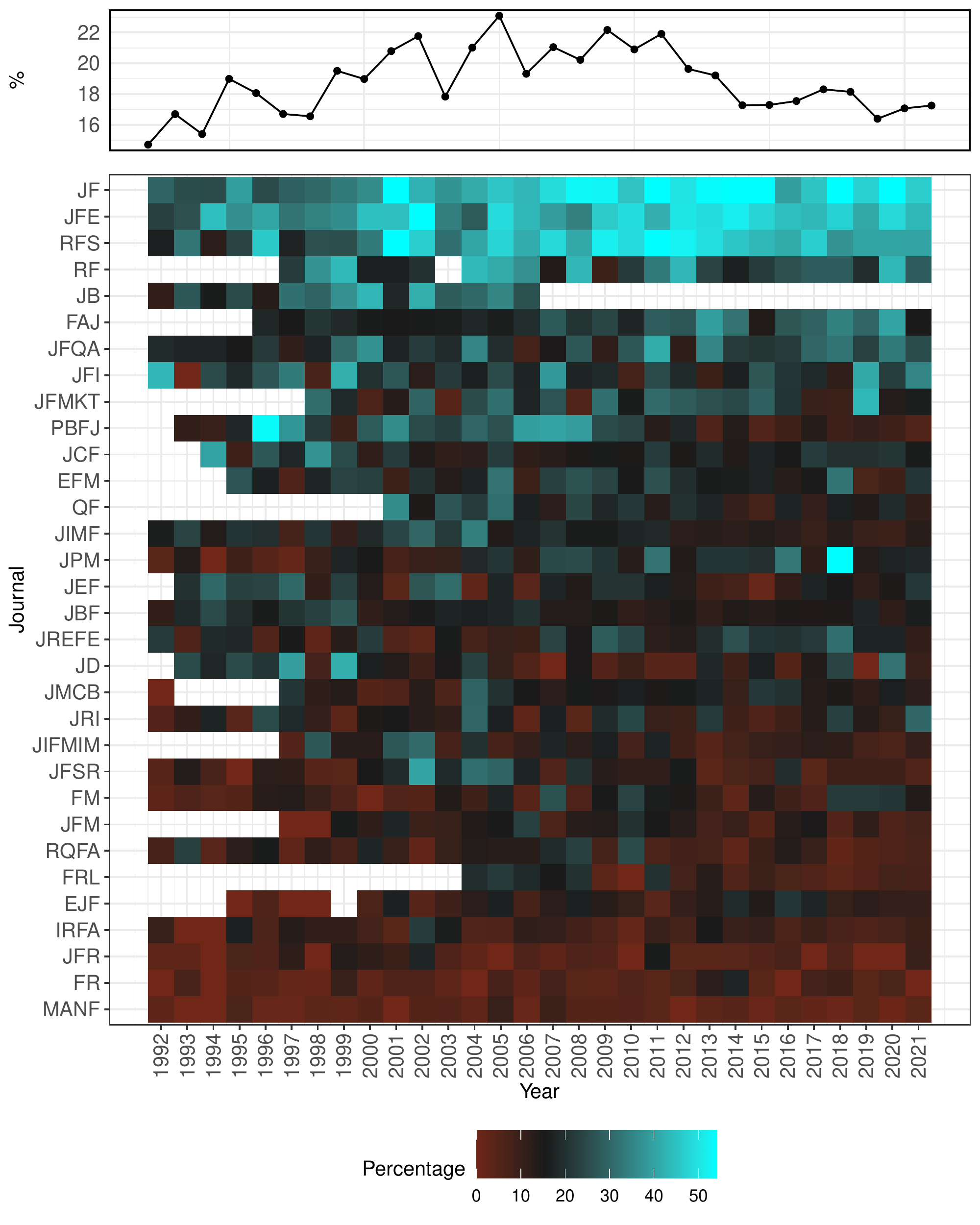}
\caption{\textbf{Proportion of Articles With at Least One Top-Tier Institution Author}\\
This figure displays the proportion of articles with at least one author affiliated with a top-tier institution in finance. The top panel shows the proportion over all 
outlets, while the bottom panel reports results per journal. Journals are sorted from higher to lower proportions computed over the 1992--2021 period.}
\label{fig:top25}
\end{figure}

In Figure \ref{fig:top25}, we display the evolution of the proportion of top-tier university articles. On average, about 15\% (in 1992) to 23\% (2005) of all articles published in our sample have at least one author with a top-tier affiliation. The largest proportions are observed for articles published in the \emph{Journal of Finance} (43\%), the \emph{Journal of Financial Economics} (43\%), and the \emph{Review of Financial Studies} (42\%). Interestingly, the practitioner-oriented \emph{Financial Analysts Journal} gained popularity among top-tier university authors. In contrast, the region-specific \emph{Pacific-Basin Finance Journal} has lost popularity since 2012.

\insertfloat{Figure~\ref{fig:top25}}

\subsection{Topicality}\label{sec:topicality}

\noindent
Our most extensive set of analyses is about the financial literature's research topics evolution, relation, and determinants. Section~\ref{sec:ntopics} provides details on the choice of the number of topics in the STM. In Section~\ref{sec:topic}, we discuss the choices of topic labels and analyze their unconditional prevalence. In Section~\ref{sec:correlation}, we measure the dependence between the topics. In Section~\ref{sec:evolution}, we investigate how topics' diversity has evolved in the field. In Section~\ref{sec:journals}, we look at topics per journal and thematic correlations across journals. Finally, in Section~\ref{sec:top}, we investigate how research topicality relates to the number of authors, gender, and top-tier institution affiliation.

\subsubsection{Optimal Number of Topics}\label{sec:ntopics}

\noindent
We use two common metrics to determine the optimal number of topics $K$ in our corpus: (i) the semantic coherence and (ii) the exclusivity. Semantic coherence is maximized when the most probable words in a given topic frequently co-occur together, and it is a metric that correlates well with the human judgment of topic quality \citep{MimnoEtAl2011}. However, having high semantic coherence is relatively easy if one has only a few topics dominated by very common words. \cite{RobertsEtAl2014} propose a complementary metric called exclusivity. It measures the exclusiveness of the words that make up a topic. The objective is to find the number of topics $K$ that balance semantic coherence and exclusivity. To do so, we calibrate the STM with $K \in \{20, 30,..., 150\}$. 

\begin{figure}[H]
\centering
\includegraphics[width=0.95\textwidth]{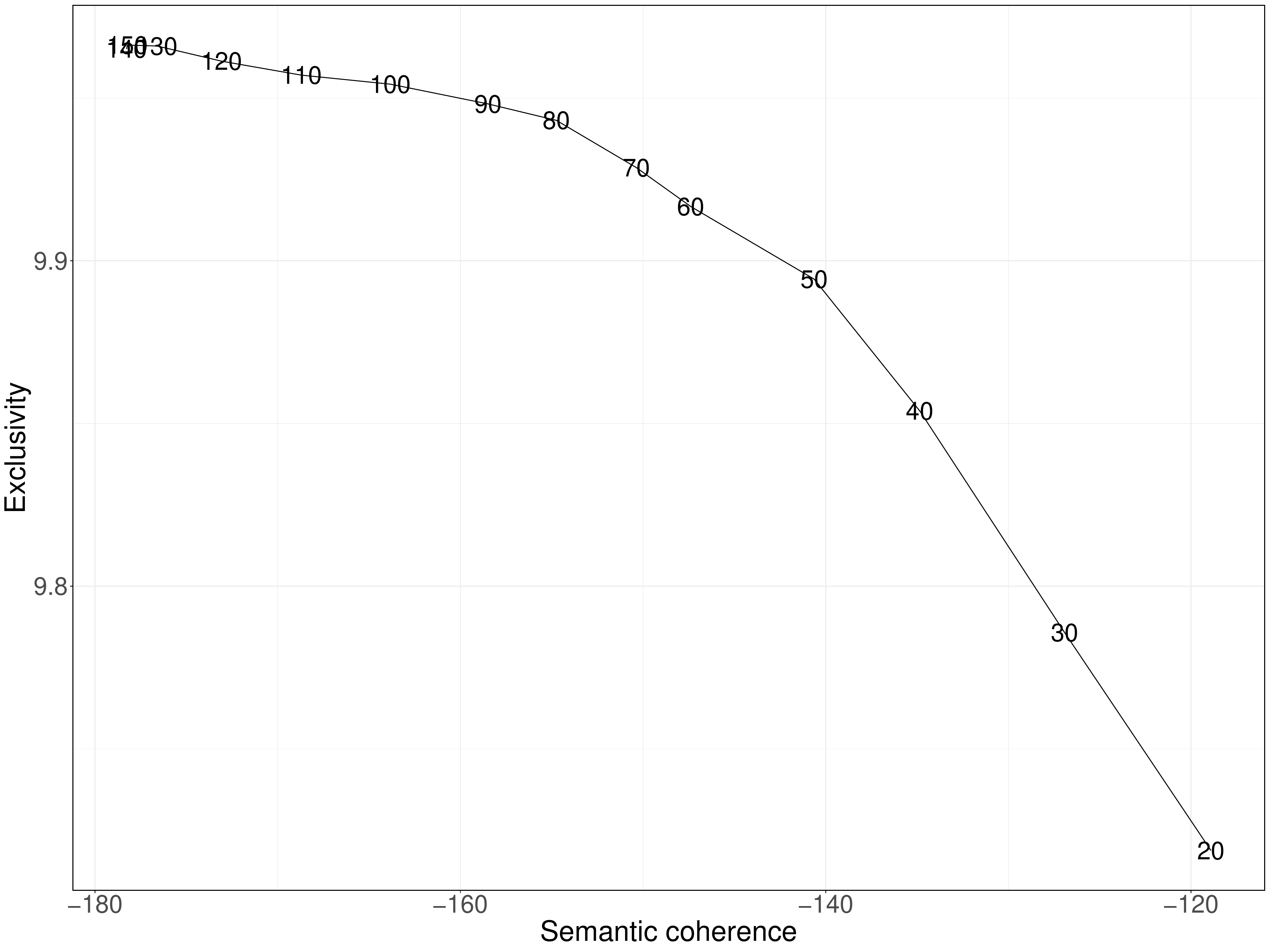}
\caption{\textbf{Semantic Coherence vs. Exclusivity}\\
This figure displays the values of semantic coherence (horizontal axis) and exclusivity (vertical axis) for various numbers of topics $K$ (from $K=20$ to $K=150$) in the STM.}
\label{fig:cohex}
\end{figure}

In Figure~\ref{fig:cohex}, we report the semantic coherence and exclusivity for each of those models. We can observe that a very high (low) number of topics leads to a high exclusivity but low (high) semantic coherence. It seems that 50 topics appears to be around the point where lowering (increasing) the number of topics would significantly reduce exclusivity (semantic coherence). Therefore, we choose $K= 50$ as the optimal number of topics for the STM estimation.

\insertfloat{Figure~\ref{fig:cohex}}

\subsubsection{Topic Labelling and Topic Prevalance}

\label{sec:topic}

\noindent
To start our analyses, we first manually label the 50 topics in our corpus. We proceed by (i) looking at the ten most probable words for each topic (ten largest $\beta_{k,v}$ values for topic $k$) and (ii) looking at the content of the articles with the largest topic prevalence (highest $\theta_{d,k}$ value for topic $k$). Let us illustrate these two steps for a topic that we name \textsf{Option Pricing}. In this case, the ten largest $\beta_{v,k}$ (expressed in percentage) are:
\begin{center}
\scalebox{0.8}{ 
\begin{tabular}{lcccccccccc}
term & \textsf{option} & \textsf{method} & \textsf{model} & \textsf{pricing} & \textsf{jump} 
& \textsf{derivative} & \textsf{price} & \textsf{formula} & \textsf{time} & \textsf{option\_price}\\ 
probability & 10.59 & 2.49 &  2.38  &2.29  &1.96  &1.74 & 1.61 & 1.53  &1.46 & 1.32
\end{tabular}}
\end{center}
From these words, it is reasonable to assume the topic is about option pricing. We can confirm this by looking at the content of the articles with a large topic prevalence: The article with the highest prevalence is ``Flexible Arithmetic Asian Options'' by \citet{zhang1995flexible}. We proceed the same way to label the other topics. In the online Appendix, Section~\ref{tab:topiclabel}, we report the ten highest-probability words for each topic in our list. Among the 50 topics, 45 can be easily associated with a particular strand of research.\footnote{Two topics are related to the analysis of the financial literature (\textsf{finance}, \textsf{uncertainty}, \textsf{research}, \textsf{article}, \textsf{project}, \textsf{literature}, \textsf{development}, \textsf{theory}, \textsf{review}, \textsf{issue}), and literature reviews (\textsf{paper}, \textsf{purpose}, \textsf{group}, \textsf{methodology}, \textsf{study}, \textsf{finding}, \textsf{approach}, \textsf{design}, \textsf{limited}, \textsf{author}). One topic relates to data analysis, which we consider too broad (\textsf{model}, \textsf{taylor}, \textsf{distribution}, \textsf{francis}, \textsf{data}, \textsf{estimation}, \textsf{approach}, \textsf{method}, \textsf{process}, \textsf{paper}). Finally, two topics have less than 1\%  of unconditional prevalence and are not clearly identifiable: (\textsf{industry}, \textsf{decision}, \textsf{profit}, \textsf{profitability}, \textsf{industries}, \textsf{participation}, \textsf{decline}, \textsf{privatization}, \textsf{operation}, \textsf{make}) and (\textsf{paper}, \textsf{result}, \textsf{effect}, \textsf{evidence}, \textsf{impact}, \textsf{analysis}, \textsf{right}, \textsf{study}, \textsf{number}, \textsf{market}).} Therefore, we keep these 45 topics for our next analyses.

In Table~\ref{tab:topic}, we report the list of topics and their unconditional prevalence across the 32 financial research journals. We find that the five most  prevalent topics are \textsf{Trading/Liquidity} (4.01\%),
\textsf{Monetary Policy} (3.29\%), \textsf{Portfolio Strategy} (3.27\%), \textsf{Option Pricing} (3.25\%), and \textsf{Market Anomaly} (3.15\%). Regarding the five least covered topics, we have \textsf{Payout Policy} (1.06\%), \textsf{Political Relationship/Corruption} (1.13\%), \textsf{Corporate Conduct/Regulation} (1.15\%), \textsf{Hedging} (1.15\%), and \textsf{Firm Valuation} (1.18\%). 

\insertfloat{Table~\ref{tab:topic}}

\subsubsection{Topic Dependence}
\label{sec:correlation}

\begin{figure}[H]
\centering
\includegraphics[width=1\textwidth]{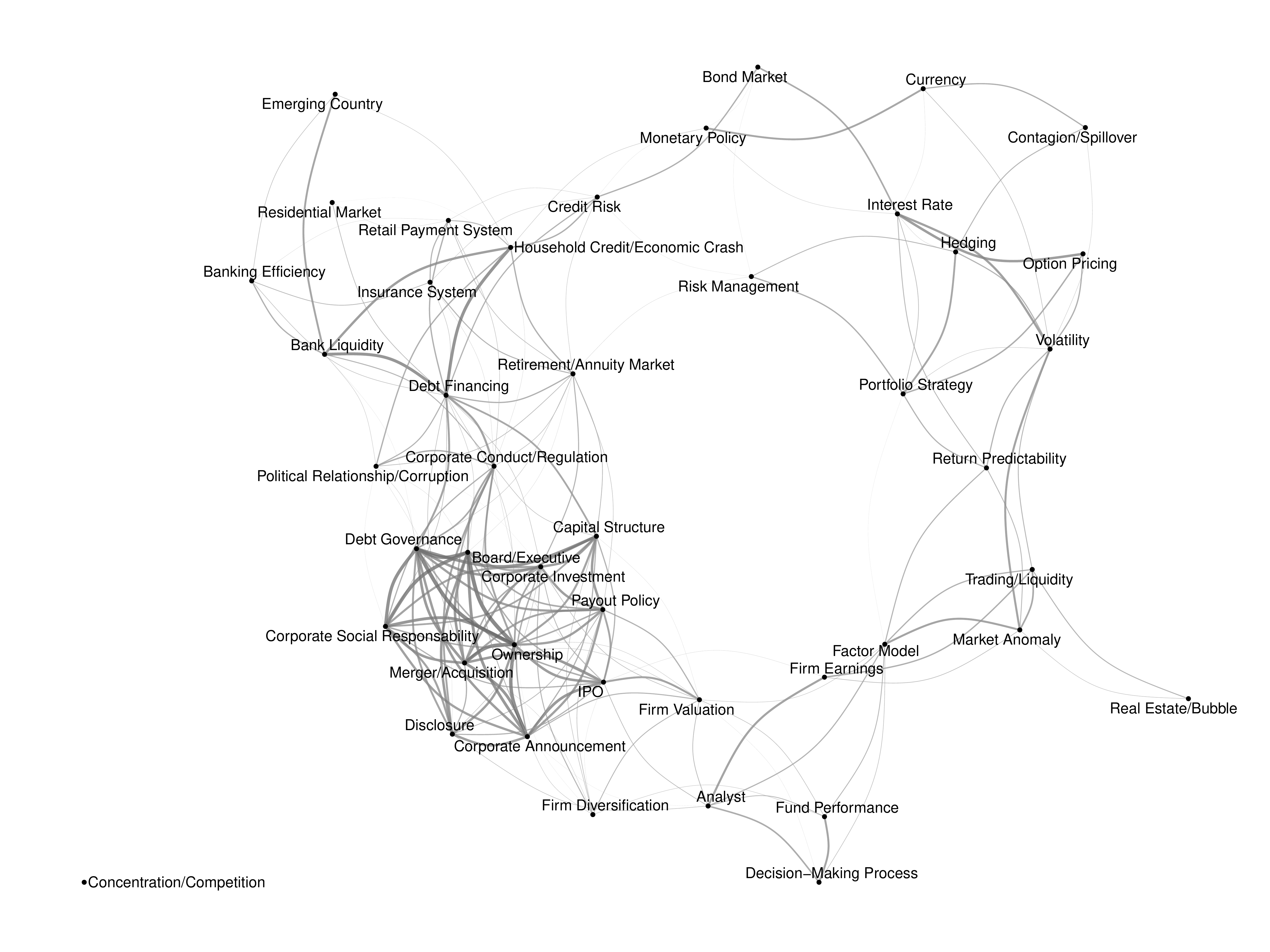}
\caption{\textbf{Topic Correlation Network}\\
This figure displays the Spearman correlation network between the 45 topics obtained from the MAP estimates of the calibrated STM; see Section~\ref{sec:testing}. 
To keep the network readable, we remove any links that are generated by a correlation below 0.45.}
\label{fig:topiccorr}
\end{figure}

\noindent
An appealing feature of the STM is that it explicitly captures the correlation among topics' prevalence. We interpret the level of correlation between topics' prevalence as the level of synergy across the subjects. The dependence structure is obtained by estimating Spearman's correlations between the $K$ columns of matrix~$\Theta$. In Figure~\ref{fig:topiccorr}, we display the correlation network of the topics' prevalence. Gray to black edges connect the topics with high correlations, where the edges are thicker (thinner) and blacker (grayer) when the correlation is higher (lower). We see strong links among some topics, which can be interpreted as clusters. Let us emphasize four clusters for the sake of illustration. 

In the bottom-left part, we notice a dense cluster of topics that can be associated with ``corporate finance'': \textsf{Corporate Announcement}, \textsf{Firm Diversification}, \textsf{IPO}, \textsf{Ownership}, \textsf{Payout Policy}, \textsf{Merger/Acquisition},  \textsf{Corporate Social Responsibility},  \textsf{Disclosure}, \textsf{Corporate Investment}, \textsf{Board/Executive}, \textsf{Capital Structure}, and \textsf{Debt Governance}. In the top-left corner, we observe topics related to ``banks and financing'' (corporate and retail): \textsf{Debt Financing}, \textsf{Retirement/Annuity Market}, \textsf{Banking Liquidity}, \textsf{Banking Efficiency}, \textsf{Insurance System}, \textsf{Retail Payment System}, \textsf{Residential Market}, and \textsf{Emerging Country}. In the top-right, we see topics related to ``financial risk'': \textsf{Option Pricing}, \textsf{Hedging}, \textsf{Volatility}, \textsf{Interest Rate}, and \textsf{Currency}. In the bottom-right, we observe topics related to ``financial market and asset pricing'': \textsf{Return Predictability},  \textsf{Trading/Liquidity},  \textsf{Market Anomaly}, \textsf{Real Estate/Bubble}, and \textsf{Factor Model}. The network is also insightful for identifying topics that bridge the clusters. For instance, \textsf{Debt Financing} links the ``corporate finance'' and ``banking and insurance'' clusters. In the top-center part, \textsf{Risk Management}, \textsf{Monetary Policy}, and \textsf{Bond Market} link ``risk management'' and ``banking and insurance.'' In the center-right part, \textsf{Portfolio Stategy} links `financial risk'' to ``financial market and asset pricing.'' Finally, in the bottom-center, \textsf{Analyst} and \textsf{Firm Valuation} link ``financial market and asset pricing'' and ``corporate finance.'' 

\insertfloat{Figure~\ref{fig:topiccorr}}

\subsubsection{Topic Evolution and Topic Concentration}
\label{sec:evolution}

\noindent
This section analyzes the highest- and lowest-prevalent topics over time. We are also interested in analyzing whether financial research has concentrated or has become broader over time. To do so, we introduce the concept of topic concentration. Topic concentration measures how much an article, or a group of articles, is concentrated in a few topics. To measure topic concentration, we use the normalized Herfindahl index. Given a vector of topic prevalence $\theta = (\theta_1,\ldots,\theta_K)$, we define topic concentration as:

\begin{equation} 
\textit{TC}(\theta) = \frac{\sum_{k=1}^K \theta_k^2 - 1/K}{1-1/K} \,.
\end{equation}

$\textit{TC}$ has a value between zero and one: Zero indicates that all $K$ topics have a prevalence of $1/K$ and one indicates concentration into a single topic in the set of documents. The yearly highest- and lowest-prevalent topics and topic concentrations are computed from the average of the topic prevalences of the articles published within a year.\footnote{Note that this approach is different than calculating the average of the highest- and lowest-prevalent topics of the articles and the average of the topic concentrations of the articles published within a year.} 

Results are reported in Table~\ref{tbl:topicyear}. First, we observe that \textsf{Market Anomaly} and \textsf{Trading/Liquidity} dominated the field in the first part of our sample. The former was the most prevalent in 1992--1994 and the latter in 1995--2005. Following 2005, \textsf{Monetary Policy} became one of the most highly researched topics among finance academics along with \textsf{Trading/Liquidity} and for one year, in 2013, \textsf{Bank Liquidity}. Because of the publication lag in academic journals, likely, both topics gained in popularity due to the 2008 Financial Crisis and the European debt crisis. A shift happened in 2017--2021 with \textsf{Corporate Social Responsibility} becoming the most research topic among all 45 topics identified in our sample. Results also highlight changes for the least covered topics. In particular, research papers related to \textsf{Political Relationship/Corruption}, \textsf{Hedging}, and \textsf{Corporate Conduct/Regulation} were absent from 1993 to 2013 relative to other topics. This result is expected as those are emerging topics in the field. These have been replaced by
\textsf{Payout Policy}, \textsf{Firm Valuation}, and \textsf{IPO} in recent years, indicating a strong decline in those research topics' popularity.

\insertfloat{Table~\ref{tbl:topicyear}}

When looking at the yearly topic concentration (multiplied by 100 for readability) in the last column of Table~\ref{tbl:topicyear}, we note a decrease over time in our sample (from 0.63 in 1992 to 0.42 in 2021). The average topic concentration was 0.57 in 1992--2006 and 0.39 in 2007--2021. A linear regression analysis indicates a significant decrease of -0.01 each year (at the 1\% significance level). This shows that the field has gained in diversity over time.

We complete the analysis above by a trend analysis to determine how topics have increased or declined in ``popularity'' over time. From the calibrated STM, we use the method of composition described in Section~\ref{sec:testing} to estimate the following $K$ beta regressions: 
\begin{align} \label{eq:trendtopic}
\begin{split}
\theta_{d,k} & \sim \textit{Beta} ( \mu_{d,k}, \phi_{k} ) \\
\mu_{d,k} & = g \left( a_{k} + b_{k}(\textit{Year}_d - 1992) \right) \,,
\end{split}
\end{align}
where $k=1,\ldots,K$, $g(\cdot)$ is the logit link function, $ \phi_{k}$ is a precision parameter, $\textit{Year}_d$ is the year of publication of the article $d$, and $a_k$ and $b_k$ are coefficients. The estimate of $\frac{\exp(a_k)}{\exp(a_k) + 1}$ measures the prevalence of topic $k$ in 1992, while the estimate of $b_k$ assesses the sensitivity of prevalence of topic $k$ with respect to time. 

We report the results in Table~\ref{tbl:evol}. We see that most topics have either decreased or increased significantly over time. We note that the topics with a decreased (increased) prevalence had a relatively large (low) prevalence in 1992. Indeed the correlation between $a_{k}$ and $ \phi_{k} $ is negative (\ie, -0.28). Thus, this results in more evenly distributed prevalences across topics over time, in line with the results above on topic concentration. The topics for which the prevalence has increased most are \textsf{Correlation/Spillover} and \textsf{Corporate Social Responsibility}. The topics the prevalence of which has decreased the most are \textsf{Market Anomaly} and \textsf{Interest Rate}. 

\insertfloat{Table~\ref{tbl:evol}}

As a complementary analysis, we evaluate whether topic prevalence's evolution has been non-linear in time, for instance, growing and then slowing down afterward. We estimate 
the following $K$ beta regressions from the calibrated STM:
\begin{align} \label{eq:trend}
\begin{split}
\theta_{d,k} & \sim \textit{Beta} ( \mu_{d,k}, \phi_{k} ) \\
\mu_{d,k} &= g\left( a_k + \sum_{y=1993}^{2021} b_{y,k} I(\textit{Year}_d = y)  \right) \,,
\end{split}
\end{align}
where $k=1,\ldots,K$, $I(\textit{Year}_d = y)$ is an indicator variable that is equal to one when the condition holds. The estimator of $b_{y,k}$ measures the sensitivity of increase in the prevalence of topic $k$ in year $y$ relative to 1992. 

\begin{figure}[H]
\centering
\includegraphics[width=1\textwidth]{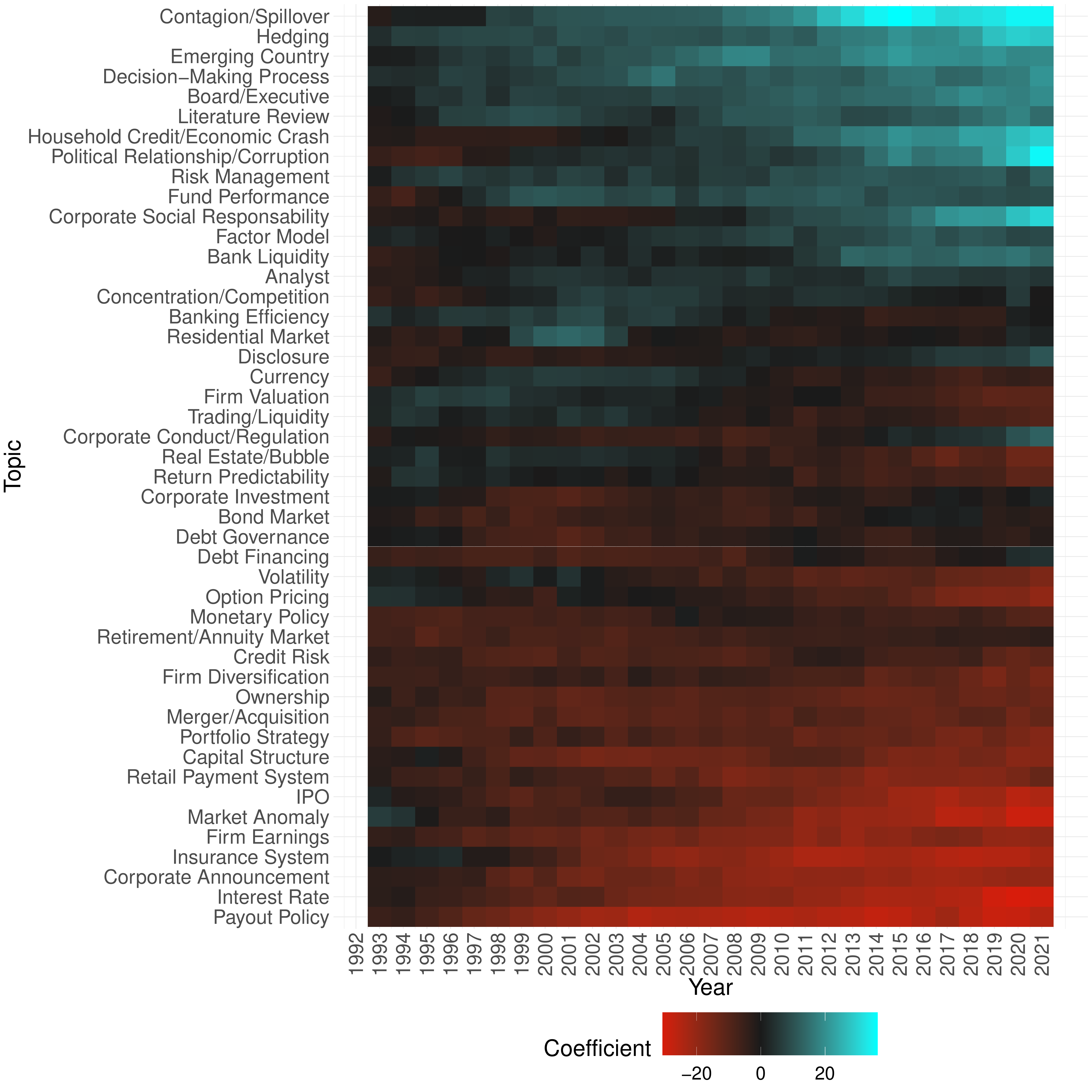}
\caption{\textbf{Topic Prevalence Over Time}\\
This figure display a heatmap of the estimates of $b_{1993,k}$ to $b_{2021,k}$ ($\times\!\!$~100) for each 
topic (vertical axis) over time (horizontal axis) 
in model~\eqref{eq:trend}. Coefficient and p-value are obtained from 
the method of composition described in Section~\ref{sec:testing}. 
Topics are sorted from higher to lower average coefficents over the 1992--2021 period.}
\label{fig:growth_topic}
\end{figure}

In Figure~\ref{fig:growth_topic}, we display a heatmap of the estimates of $b_{1993,k}$ to $b_{2021,k}$ for each topic (vertical axis) over time (horizontal axis). We see the sharp increase in prevalence for \textsf{Correlation/Spillover} around 2009, near the financial crisis, and since 2016 for \textsf{Corporate Social Responsibility}. Interestingly, \textsf{Residential Market} exhibits a high level of prevalence between 1998 and 2002, but not in other years. Similarly, \textsf{Fund Performance} increased strongly in 1998 but has exhibited a decline since then. Overall, this illustrates how dynamic academic finance is regarding the subjects covered over the last 30 years.

\insertfloat{Figure~\ref{fig:growth_topic}}

\subsubsection{Topics and Journals}
\label{sec:journals}

\noindent
We now turn to topic analyses conditional on journals. In Table~\ref{tbl:journal_topic}, we report the highest- and lowest-prevalent topics together with the topic concentration for each journal in our sample. The quantities are obtained from the average topic prevalences of the articles published by each journal.

First, we see that several journals are heavily specialized (\ie, high TC), some with a highest-prevalent topic above 20\%. The most specialized journals are the \emph{Journal of Derivatives} (TC at 15.10, \textsf{Option Pricing} at 34.11\%), the \emph{Journal of Risk and Insurance} (TC at 10.98, \textsf{Insurance System} at 29.13\%), and the \emph{Journal of Money, Credit and Banking} (TC at 10.88, \textsf{Monetary Policy} at 30.60\%). On the other hand, the most generalist journals are the \emph{European Journal of Finance} (TC at 0.53), \emph{European Financial Management} (TC at 0.54), and \emph{Managerial Finance} (TC at 0.56). We also note that the \emph{Review of Financial Studies}, the \emph{Journal of Financial Economics}, the \emph{Journal of Finance}, the \emph{Journal of Financial and Quantitative Analysis}, and the \emph{Review of Finance} are also close to these outlets in terms of concentration. Thus, commonly recognized top-five journals in finance cover a broad set of topics. With a topic concentration at 0.68, the \emph{Journal of Banking \& Finance} can also be considered a generalist outlet.

\insertfloat{Table~\ref{tbl:journal_topic}}

To investigate how journals' coverages have evolved over time, we look at the sensitivity of the topic concentrations with respect to time for each outlet. We estimate the following $J$ beta regressions from the calibrated STM:
\begin{align} \label{eq:trendtc}
\begin{split}
\textit{TC}_{y,j}     
& \sim \textit{Beta} ( \mu_{y,j}, \phi_j ) \\
\mu_{y,j}  
& = g \left( a_j + b_j  \Delta\!\textit{Year}_{y,j} \right) \,,
\end{split}
\end{align}
where $j=1,\ldots,J$ denotes the journal, $\textit{TC}_{y,j} = \textit{TC}(\theta_{y,j})$ is the topic concentration of journal $j$ in year $y$ (computed as the average topic prevalences), $\Delta\!\textit{Year}_{y,j}$ is the difference between year $y$ and the first year of publication of outlet $j$, and $a_j$ and $b_j$ are journal-specific level and trend in topic concentrations. We are specifically interested in the estimates of $b_j$. 

Results are reported in the last column of Table~\ref{tbl:journal_topic}. We find that 24 journals have experienced a decrease in their topic concentration over time. Among them, 15 show a significant negative coefficient $\beta_j$ at the 1\% level (and 17 journals a negative coefficient at the 10\% level).
On the contrary, eight journals show an increase in concentration, but only the \emph{Financial Analysts Journal} exhibits a significant coefficient exposure to time at the 1\% significance level. Overall, these results support the previous findings that the field of academic finance has become broader over time.

\begin{figure}[H]
\centering
\includegraphics[width=1\textwidth]{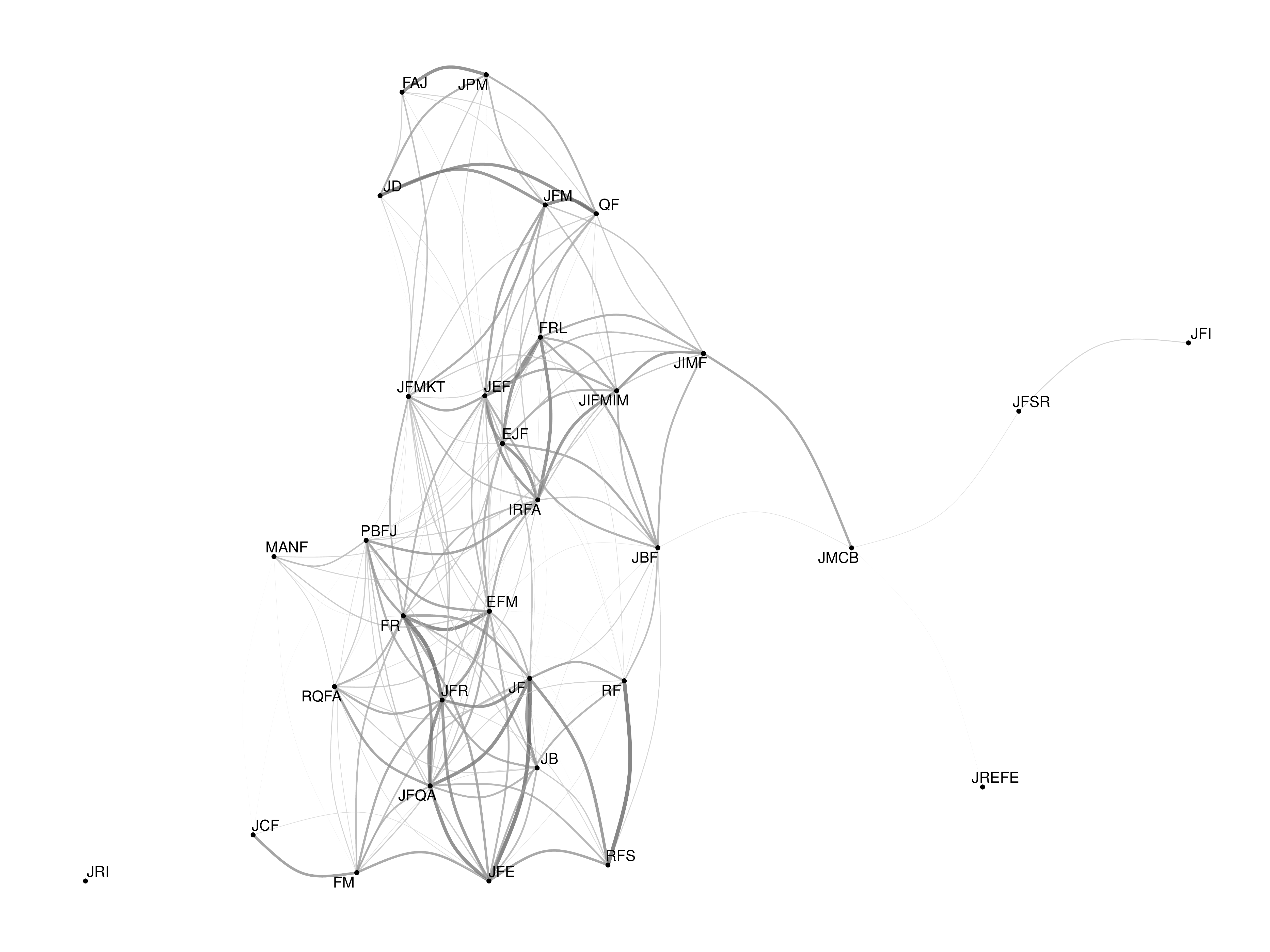}
\caption{\textbf{Journal Correlation Network}\\
This figure displays the Spearman correlation network between the topics discussed in 
the journals (\ie, between the average topic prevalence obtained from the MAP estimates of the calibrated STM; see Section~\ref{sec:testing}).
To keep the network readable, we remove any links that are generated by a correlation below 0.45.}
\label{fig:journal_rel}
\end{figure}

To investigate topical relationships between journals, we display the correlation network of the journal-specific topic prevalences in Figure~\ref{fig:journal_rel}. We obtain this network by computing the correlation of the journals' average topic prevalences. From the network, we note the close relationship between the \emph{Journal of Financial Economics}, the \emph{Journal of Finance}, the \emph{Journal of Financial and Quantitative Analysis}, the \emph{Review of Financial Studies}, and the \emph{Journal of Financial Research}. The \emph{Journal of Financial Research} is also well connected to the \emph{Financial Review}. The \emph{Journal of Corporate Finance} is only well connected to \emph{Financial Management}, which is well connected to the \emph{Journal of Financial Economics} and the \emph{Journal of Financial and Quantitative Analysis}. We also observe a strong connection between \emph{Finance Research Letters}, the \emph{International Review of Financial Analysis}, the \emph{European Journal of Finance}, the \emph{Journal of Empirical Finance}, and the \emph{Journal of International Financial Markets Institutions \& Money}. Somewhat well connected to that cluster is also the \emph{Journal of Banking \& Finance}. \emph{Quantitative Finance} is connected to the \emph{Journal of Futures Markets} and the \emph{Journal of Derivatives}. In addition, the two practitioner-oriented outlets, the \emph{Journal of Portfolio Management} and the \emph{Financial Analysts Journal}, are strongly connected. Finally, we see that that the \emph{Journal of Risk and Insurance} is set apart and does not exhibit strong connections with other outlets. As such, this journal predominantly covers topics not covered by other journals.

\insertfloat{Figure~\ref{fig:journal_rel}}

Finally, we complete the outlets analysis by looking at the topic prevalence that has most increased or decreased over time per journal. We estimate the following $K$ beta regressions from the calibrated STM:
\begin{align} \label{eq:trendtopic2}
\begin{split}
\theta_{d,k} & \sim \textit{Beta} ( \mu_{d,k}, \phi_{k} ) \\
\mu_{d,k}  & = g \left(  \sum_{j=1}^{32} I(\textit{Jrnl}_d = j)  \left( a_{j,k} 
+ b_{j,k} \Delta\!\textit{Year}_{d} \right) \right) \,,
\end{split}
\end{align}
where $k=1,\ldots,K$, $I(\textit{Jrnl}_d = j)$ is an indicator function the value of which is one when article $d$ is published in journal $j$ and 
$ \Delta\!\textit{Year}_{d}$ is the difference between the year of the article's publication and the first year in our database of the journal in which the article appeared. 

In Table~\ref{tbl:journal_topic_evol}, we report, for each journal, the two highest and lowest estimates of $b_{j,k}$. While we see heterogeneity among the outlets regarding increasing or decreasing topic prevalences, we see commonalities. The most often seen growing topics are \textsf{Household Credit/Economic Crash} (20 occurrences) and \textsf{Corporate Social Responsibility} (14 occurrences). In comparison, the most often seen decreasing topic is \textsf{Market Anomaly} (16 occurrences).

\insertfloat{Table~\ref{tbl:journal_topic_evol}}

\subsection{Topics and Research Team Characteristics}
\label{sec:top}

\noindent
We complete our analysis by investigating how topic prevalences are related to the following research team characteristics: (i) the number of authors, (ii) at least one female in the team, and (iii) at least one author in the team affiliated with a top-tier institution in finance. We estimate the following $K$ beta regressions from the calibrated STM:
\begin{align} \label{eq:topuniv}
\begin{split}
\theta_{d,k} & \sim \textit{Beta} ( \mu_{d,k}, \phi_{k} ) \\
\mu_{d,k} & = g \left(  a_{k} 
+ b_{k} \Delta\!\textit{Year}_d 
+ c_{k} \textit{NAuth}_d
+ d_{k} I(\textit{Woman}_d = 1) 
+ e_{k} I(\textit{Top}_d = 1)   \right) \,,
\end{split}
\end{align}
where $k=1,\ldots,K$, $\textit{NAuth}_d$ is the number of authors of articles $d$, $I(\textit{Woman}_d = 1)$ is a dummy variable the value of which is one if at least one author of article $d$ is a female, and $I(\textit{Top}_d = 1)$ is a dummy variable the value of which is one if at least one author of article $d$ is affiliated to a top-tier institution in the finance. 

In Table~\ref{tbl:team}, we report the top six positive (left) and negative (right) topic prevalence sensitivity coefficients in model \eqref{eq:topuniv} for the research's team characteristics.
In Panel~A, we find that larger teams tend to publish more about \textsf{Corporate Announcement}, \textsf{Market Anomaly}, and \textsf{Disclosure}, and less about \textsf{Monetary Policy}, \textsf{Retail Payment System}, and \textsf{Interest Rate} as compared to smaller teams. In Panel~B, we find that teams with at least one female tend to publish more about \textsf{Corporate Social Responsibility}, \textsf{Board/Executive}, and \textsf{Ownership} as compared to male-only teams. On the contrary, we find a negative sensitivity of topic prevalence for \textsf{Return Predictability}, \textsf{Portfolio Strategy}, or \textsf{Volatility}. Interestingly, the first set of topics is related to social and governance aspects of ``corporate finance'', while the latter is oriented towards technical and modeling aspects. Finally, in Panel~C, we find that teams with at least one top-tier institution scholar tend to focus more on \textsf{Decision-Making Process}, \textsf{Corporate Investment}, and \textsf{Debt Financing}, and less on \textsf{Contagion/Spillover}, \textsf{Emerging Country}, and \textsf{Volatility} as compared to other teams. The first three topics seem to cover more theoretical aspects of finance, while the three others are rather empirical. This conclusion 
is supported when looking at the MAP estimates of the word probability. When we compute the sum of the probabilities of the set of vocabulary keywords: ``equilibrium'', ``preference'', ``theory'', and ``equilibrium\_model'', minus the sum of the probabilities of the set of vocabulary keywords: ``empirical\_analysis'', ``empirical\_result'', ``empirical\_evidence'' for each topic, we find that the former three topics rank higher (positions 1, 10, and 13) than the latter three topics (positions 39, 44, and 45).

\insertfloat{Table~\ref{tbl:team}}

\section{Conclusion}
\label{sec:conclusion}

\noindent
We study how the financial literature has evolved in scale, research team composition, and article topicality across 32 finance-focused academic journals from 1992 to 2021. We document that the field has vastly expanded regarding outlets and published articles. Teams have become larger, and the proportion of women participating in research has increased significantly. Using the Structural Topic Model, we show that the landscape of academic finance can be represented by 45 connected topics that can be regrouped 
into four major themes: ``corporate finance'', ``banking and insurance'', ``financial risk'', and ``financial market and asset pricing''. We investigate the topic coverage of individual journals and can identify highly specialized and more generalist outlets, but our analyses reveal that most journals have covered more topics over time, thus becoming more generalist. Finally, we find that articles with at least one woman author focus more on topics related to social and governance aspects of ``corporate finance''. On the contrary, the prevalence is less pronounced for topics oriented toward technical and modeling aspects in finance. We also find that teams with at least one top-tier institution scholar tend to focus more on theoretical aspects of finance.

We believe our landscape will help scholars to better position their research in the field and to find relevant journals to submit their work. Alike \citet{CorbetEtAl2019}, we also hope our study will provide editors and conference organizers with ways to identify gaps in the field, leading them to create special issues or conference sessions about topics that are generally not co-occurring, thus creating potential new synergies in the domain.


\newpage
\singlespacing
\bibliography{ref}

\newpage
\begin{table}[H]
\caption{\textbf{List of Academic Journals}\\
The table reports the name, the abbreviation, the total number of articles published, the yearly average 
of published articles, and the starting and ending date in our sample of each 32 journals in our study.}
\label{tab:journal}
\centering
\scalebox{0.76}{
\begin{tabular}{llrccc}
\toprule
Journal & Abbr. & Total  & Average & Start & End \\ 
\midrule
European Financial Management & EFM & 809 & 30 & 1995 & 2021 \\ 
European Journal of Finance & EJF & 1,172 & 45 & 1995 & 2021 \\ 
Financial Analysts Journal & FAJ & 859 & 33 & 1996 & 2021 \\ 
Financial Management & FM & 905 & 30 & 1992 & 2021 \\ 
Financial Review & FR & 916 & 31 & 1,992 & 2021 \\ 
Finance Research Letters & FRL & 1,693 & 94 & 2004 & 2021 \\ 
International Review of Financial Analysis & IRFA & 1,819 & 61 & 1992 & 2021 \\ 
Journal of Business & JB & 461 & 31 & 1,992 & 2006 \\ 
Journal of Banking \& Finance & JBF & 4,884 & 163 & 1992 & 2021 \\ 
Journal of Corporate Finance & JCF & 1,935 & 69 & 1994 & 2021 \\ 
Journal of Derivatives & JD & 581 & 20 & 1993 & 2021 \\ 
Journal of Empirical Finance & JEF & 1,221 & 42 & 1993 & 2021 \\ 
Journal of Finance & JF & 2,330 & 78 & 1,992 & 2021 \\ 
Journal of Financial Economics & JFE & 2,750 & 92 & 1992 & 2021 \\ 
Journal of Financial Intermediation & JFI & 619 & 21 & 1992 & 2021 \\ 
Journal of Futures Markets & JFM & 1,324 & 53 & 1997 & 2021 \\ 
Journal of Financial Markets & JFMKT & 588 & 24 & 1998 & 2021 \\ 
Journal of Financial and Quantitative Analysis & JFQA & 1,484 & 49 & 1992 & 2021 \\ 
Journal of Financial Research & JFR & 807 & 27 & 1,992 & 2021 \\ 
Journal of Financial Services Research & JFSR & 636 & 21 & 1992 & 2021 \\ 
Journal of International Financial Markets Institutions \& Money & JIFMIM & 1,324 & 53 & 1997 & 2021 \\ 
Journal of International Money and Finance & JIMF & 2,296 & 77 & 1992 & 2021 \\ 
Journal of Money Credit and Banking & JMCB & 1,666 & 64 & 1992 & 2021 \\ 
Journal of Portfolio Management & JPM & 1,355 & 45 & 1992 & 2021 \\ 
Journal of Real Estate Finance and Economics & JREFE & 1,274 & 42 & 1992 & 2021 \\ 
Journal of Risk and Insurance & JRI & 964 & 32 & 1992 & 2021 \\ 
Managerial Finance & MANF & 1,882 & 63 & 1992 & 2021 \\ 
Pacific-Basin Finance Journal & PBFJ & 1,479 & 51 & 1993 & 2021 \\ 
Quantitative Finance & QF & 1,786 & 85 & 2001 & 2021 \\ 
Review of Finance & RF & 729 & 30 & 1997 & 2021 \\ 
Review of Financial Studies & RFS & 2,108 & 70 & 1992 & 2021 \\ 
Review of Quantitative Finance and Accounting & RQFA & 1,488 & 50 & 1992 & 2021 \\ 
\midrule
\multicolumn{2}{r}{Minimum} & 461  & 20 & &  \\
\multicolumn{2}{r}{Maximum} & 4,884 & 163  & & \\
\multicolumn{2}{r}{Median} & 1,324 & 47 & & \\
\multicolumn{2}{r}{Average} & 1,442 & 52  & & \\
\multicolumn{2}{r}{Total} & 46,144 & & &  \\
\bottomrule
\end{tabular}}
\end{table}

\newpage
\begin{table}[H]
\caption{\textbf{List of Topics and Topics' Unconditional Prevalence}\\
This table reports the 45 topics identified in our corpus along with their unconditional prevalence 
obtained from the MAP estimates of the calibrated STM; see Section~\ref{sec:testing}.}
\label{tab:topic}
\centering
\scalebox{0.95}{
\begin{tabular}{lc}
\toprule
Topic & \% \\
\midrule
\textsf{Trading/Liquidity} & 4.01 \\ 
\textsf{Monetary Policy} & 3.29 \\ 
\textsf{Portfolio Strategy} & 3.27 \\ 
\textsf{Option Pricing} & 3.25 \\ 
\textsf{Market Anomaly} & 3.15 \\ 
\textsf{Factor Model} & 3.06 \\ 
\textsf{Return Predictability} & 3.01 \\ 
\textsf{Bank Liquidity} & 2.99 \\ 
\textsf{Corporate Social Responsibility} & 2.81 \\ 
\textsf{Contagion/Spillover} & 2.51 \\ 
\textsf{Emerging Country} & 2.48 \\ 
\textsf{Real Estate/Bubble} & 2.41 \\ 
\textsf{Volatility} & 2.39 \\ 
\textsf{Firm Earnings} & 2.37 \\ 
\textsf{Currency} & 2.36 \\ 
\textsf{Fund Performance} & 2.34 \\ 
\textsf{Board/Executive} & 2.31 \\ 
\textsf{Decision-Making Process} & 2.11 \\ 
\textsf{Corporate Investment} & 2.08 \\ 
\textsf{Interest Rate} & 1.88 \\ 
\textsf{Debt Governance} & 1.77 \\ 
\textsf{Household Credit/Economic Crash} & 1.73 \\ 
\textsf{Risk Management} & 1.71 \\ 
\textsf{Residential Market} & 1.67 \\ 
\textsf{Ownership} & 1.67 \\ 
\textsf{Insurance System} & 1.60 \\ 
\textsf{Debt Financing} & 1.60 \\ 
\textsf{IPO} & 1.57 \\ 
\textsf{Retail Payment System} & 1.56 \\ 
\textsf{Corporate Announcement} & 1.48 \\ 
\textsf{Merger/Acquisition} & 1.47 \\ 
\textsf{Credit Risk} & 1.46 \\ 
\textsf{Disclosure} & 1.40 \\ 
\textsf{Capital Structure} & 1.38 \\ 
\textsf{Bond Market} & 1.35 \\ 
\textsf{Banking Efficiency} & 1.31 \\ 
\textsf{Retirement/Annuity Market} & 1.28 \\ 
\textsf{Analyst} & 1.24 \\ 
\textsf{Concentration/Competition} & 1.18 \\ 
\textsf{Firm Diversification} & 1.18 \\ 
\textsf{Firm Valuation} & 1.15 \\ 
\textsf{Hedging} & 1.15 \\ 
\textsf{Corporate Conduct/Regulation} & 1.15 \\ 
\textsf{Political Relationship/Corruption} & 1.13 \\ 
\textsf{Payout Policy} & 1.06 \\ 
\bottomrule
\end{tabular}}
\end{table}

\newpage
\begin{table}[H] 
\centering 
\caption{\textbf{Highest- and Lowest-Prevalent Topics and Topic Concentration Over Time}\\
The table reports for each year the highest- and lowest-prevalent topics obtained from the MAP estimates of the calibrated STM; see Section~\ref{sec:testing}. The last column reports the topic
concentration ($\times\!\!$~100) of all articles published that year. The yearly highest- and lowest-prevalent topics and topic concentrations are computed from the average of the topic prevalences of the articles published within a year.} 
\label{tbl:topicyear}
\centering
\scalebox{0.9}{ 
\begin{tabular}{ccccccc} 
\toprule
& \multicolumn{2}{c}{Highest Prevalence} & \multicolumn{2}{c}{Lowest Prevalence}  & \\ 
\cmidrule(lr){2-3}\cmidrule(lr){4-5}
Year & Topic & \% & Topic & \% & TC \\ 
\midrule
1992 & \textsf{Market Anomaly} & 4.93 & \textsf{Hedging} & 0.40 & 0.63 \\ 
1993 & \textsf{Market Anomaly} & 5.87 & \textsf{Political Relationship/Corruption} & 0.41 & 0.67 \\ 
1994 & \textsf{Market Anomaly} & 5.87 & \textsf{Political Relationship/Corruption} & 0.35 & 0.71 \\ 
1995 & \textsf{Trading/Liquidity} & 4.90 & \textsf{Political Relationship/Corruption} & 0.31 & 0.63 \\ 
1996 & \textsf{Trading/Liquidity} & 4.74 & \textsf{Political Relationship/Corruption} & 0.33 & 0.53 \\ 
1997 & \textsf{Trading/Liquidity} & 5.07 & \textsf{Political Relationship/Corruption} & 0.45 & 0.54 \\ 
1998 & \textsf{Trading/Liquidity} & 5.62 & \textsf{Political Relationship/Corruption} & 0.42 & 0.59 \\ 
1999 & \textsf{Trading/Liquidity} & 5.16 & \textsf{Political Relationship/Corruption} & 0.58 & 0.59 \\ 
2000 & \textsf{Trading/Liquidity} & 5.18 & \textsf{Political Relationship/Corruption} & 0.55 & 0.50 \\ 
2001 & \textsf{Trading/Liquidity} & 5.22 & \textsf{Political Relationship/Corruption} & 0.60 & 0.57 \\ 
2002 & \textsf{Trading/Liquidity} & 5.14 & \textsf{Political Relationship/Corruption} & 0.63 & 0.56 \\ 
2003 & \textsf{Trading/Liquidity} & 5.66 & \textsf{Political Relationship/Corruption} & 0.58 & 0.57 \\ 
2004 & \textsf{Trading/Liquidity} & 5.11 & \textsf{Political Relationship/Corruption} & 0.57 & 0.52 \\ 
2005 & \textsf{Trading/Liquidity} & 5.09 & \textsf{Political Relationship/Corruption} & 0.65 & 0.48 \\ 
2006 & \textsf{Monetary Policy} & 4.57 & \textsf{Political Relationship/Corruption} & 0.63 & 0.50 \\ 
2007 & \textsf{Trading/Liquidity} & 4.04 & \textsf{Hedging} & 0.75 & 0.40 \\ 
2008 & \textsf{Monetary Policy} & 4.29 & \textsf{Corporate Conduct/Regulation} & 0.79 & 0.45 \\ 
2009 & \textsf{Trading/Liquidity} & 4.36 & \textsf{Hedging} & 0.80 & 0.43 \\ 
2010 & \textsf{Monetary Policy} & 4.20 & \textsf{Political Relationship/Corruption} & 0.80 & 0.40 \\ 
2011 & \textsf{Monetary Policy} & 3.88 & \textsf{Hedging} & 0.75 & 0.33 \\ 
2012 & \textsf{Monetary Policy} & 3.71 & \textsf{Payout Policy} & 0.96 & 0.33 \\ 
2013 & \textsf{Bank Liquidity} & 3.89 & \textsf{Corporate Conduct/Regulation} & 0.97 & 0.38 \\ 
2014 & \textsf{Trading/Liquidity} & 4.25 & \textsf{Payout Policy} & 0.81 & 0.41 \\ 
2015 & \textsf{Trading/Liquidity} & 3.90 & \textsf{IPO} & 0.92 & 0.38 \\ 
2016 & \textsf{Factor Model} & 3.75 & \textsf{Firm Valuation} & 0.95 & 0.38 \\ 
2017 & \textsf{Corporate Social Responsibility} & 3.82 & \textsf{Firm Valuation} & 0.97 & 0.36 \\ 
2018 & \textsf{Corporate Social Responsibility} & 3.86 & \textsf{Firm Valuation} & 0.89 & 0.41 \\ 
2019 & \textsf{Corporate Social Responsibility} & 3.87 & \textsf{Payout Policy} & 0.78 & 0.39 \\ 
2020 & \textsf{Corporate Social Responsibility} & 4.60 & \textsf{Payout Policy} & 0.82 & 0.42 \\ 
2021 & \textsf{Corporate Social Responsibility} & 4.75 & \textsf{Firm Valuation} & 0.86 & 0.42 \\ 
\bottomrule
\end{tabular}} 
\end{table} 

\newpage
\begin{table}[H] 
\centering 
\caption{\textbf{Time-Trends in Topic Prevalence}\\ 
This table reports the unconditional prevalence in 1992 and time-trend prevalence for the 45 topics 
covered in our study. The method of composition described in Section~\ref{sec:testing} is used to obtain the estimates and p-values. Constant is the estimate of $\frac{\exp (a_k)}{\exp (a_k) + 1}$ and the time-trend is the estimate of $b_k$ in 
model~\eqref{eq:trendtopic}. Coefficients are multiplied by 100 and results are sorted from the highest to the lowest time-trend value. 
Signs $^{***}$, $^{**}$, and $^{*}$ indicate that the coefficients are significantly different from zero at the 1\%, 5\%, and 10\% levels.}
\label{tbl:evol}
\scalebox{0.85}{ 
\begin{tabular}{lcc} 
\toprule
Topic & Constant & Time-Trend\\
\midrule
\textsf{Contagion/Spillover} & \grbbb{2.33} & \grbbb{1.42} \\ 
\textsf{Corporate Social Responsibility} & \grbbb{2.23} & \grbbb{1.27} \\ 
\textsf{Household Credit/Economic Crash} & \grbbb{1.71} & \grbbb{1.27} \\ 
\textsf{Political Relationship/Corruption} & \grbbb{1.28} & \grbbb{1.23} \\ 
\textsf{Hedging} & \grbbb{1.42} & \grbbb{0.86} \\ 
\textsf{Bank Liquidity} & \grbbb{2.87} & \grbbb{0.70} \\ 
\textsf{Emerging Country} & \grbbb{2.62} & \grbbb{0.65} \\ 
\textsf{Board/Executive} & \grbbb{2.70} & \grbbb{0.63} \\ 
\textsf{Corporate Conduct/Regulations} & \grbbb{1.47} & \grbbb{0.5} \\ 
\textsf{Decision-Making Process} & \grbbb{2.02} & \grbbb{0.49} \\ 
\textsf{Disclosure} & \grbbb{1.64} & \grbbb{0.48} \\ 
\textsf{Factor Model} & \grbbb{2.8} & \grbbb{0.41} \\ 
\textsf{Debt Financing} & \grbbb{2.00} & \grbbb{0.35} \\ 
\textsf{Risk Management} & \grbbb{1.61} & \grbbb{0.30} \\ 
\textsf{Fund Performance} & \grbbb{2.73} & \grbbb{0.27} \\ 
\textsf{Analyst} & \grbbb{1.65} & \grbbb{0.19} \\ 
\textsf{Corporate Investment} & \grbbb{2.09} & \grbbb{0.19} \\ 
\textsf{Bond Market} & \grbbb{1.73} & \grbbb{0.18} \\ 
\textsf{Retirement/Annuity Market} & \grbbb{1.74} & \grbb{0.16} \\ 
\textsf{Debt Governance} & \grbbb{2.26} &  0.06 \\ 
\textsf{Concentration/Competition} & \grbbb{1.57} &  0.03 \\ 
\textsf{Monetary Policy} & \grbbb{3.93} & -0.09 \\ 
\textsf{Residential Market} & \grbbb{2.23} & -0.10 \\ 
\textsf{Credit Risk} & \grbbb{1.99} & \grb{-0.11} \\ 
\textsf{Merger/Acquisition} & \grbbb{2.15} & \grbbb{-0.2} \\ 
\textsf{Ownership} & \grbbb{2.17} & \grbbb{-0.28} \\ 
\textsf{Banking Efficiency} & \grbbb{1.91} & \grbbb{-0.36} \\ 
\textsf{Currencies} & \grbbb{2.92} & \grbbb{-0.39} \\ 
\textsf{Portfolio Strategy} & \grbbb{3.66} & \grbbb{-0.39} \\ 
\textsf{Capital Structure} & \grbbb{2.12} & \grbbb{-0.41} \\ 
\textsf{Firm Diversification} & \grbbb{1.43} & \grbbb{-0.42} \\ 
\textsf{Firm Earnings} & \grbbb{2.66} & \grbbb{-0.46} \\ 
\textsf{Retail Payment System} & \grbbb{2.09} & \grbbb{-0.47} \\ 
\textsf{Return Predictability} & \grbbb{3.67} & \grbbb{-0.48} \\ 
\textsf{Trading/Liquidity} & \grbbb{4.58} & \grbbb{-0.49} \\ 
\textsf{Payout Policy} & \grbbb{1.77} & \grbbb{-0.58} \\ 
\textsf{Corporate Announcements} & \grbbb{2.11} & \grbbb{-0.59} \\ 
\textsf{Firm Valuation} & \grbbb{1.49} & \grbbb{-0.63} \\ 
\textsf{Volatility} & \grbbb{3.14} & \grbbb{-0.66} \\ 
\textsf{Real Estate/Bubble} & \grbbb{2.6} & \grbbb{-0.67} \\ 
\textsf{Option Pricing} & \grbbb{4.61} & \grbbb{-0.78} \\ 
\textsf{IPO} & \grbbb{2.56} & \grbbb{-0.88} \\ 
\textsf{Insurance System} & \grbbb{2.43} & \grbbb{-0.91} \\ 
\textsf{Interest Rate} & \grbbb{2.85} & \grbbb{-0.95} \\ 
\textsf{Market Anomaly} & \grbbb{3.89} & \grbbb{-1.06} \\ 
\bottomrule
\end{tabular}}
\end{table} 

\newpage
\begin{table}[H] 
\centering 
\caption{\textbf{Highest- and Lowest-Prevalent Topics and Topic Concentration per Journal}\\
The table reports, for each journal, the highest- and lowest-prevalent topics.
TC reports the topic concentration ($\times\!\!$~100) of the journal. The quantities are obtained from the average topic prevalences of the 
articles published by each journal, where each article's topic prevalence is the MAP estimate of the calibrated STM; see Section~\ref{sec:testing}. $b_j$ is the journal-specific time-trend
sensitivity in model~\eqref{eq:trendtc} ($\times\!\!$~100).  Coefficient and p-value are obtained from 
the method of composition described in Section~\ref{sec:testing}. 
Signs $^{***}$, $^{**}$, and $^{*}$ indicate that the coefficients are significantly 
different from zero at the 1\%, 5\%, and 10\% levels.} 
\label{tbl:journal_topic}
\scalebox{0.77}{ 
\begin{tabular}{llrlrrc} 
\toprule
& \multicolumn{2}{c}{Highest Prevalence} 
& \multicolumn{2}{c}{Lowest Prevalence}  
& &\\
\cmidrule(lr){2-3}\cmidrule(lr){4-5}
Journal & Topic & \% & Topic & \% & TC & $b_j$ \\ 
\midrule
EFM & \textsf{Currency} & 4.93 & \textsf{Residential Market} & 0.47 & 0.54 & \grbbb{-3.76} \\ 
EJF & \textsf{Market Anomaly} & 4.01 & \textsf{Residential Market} & 0.45 & 0.53 & \grbbb{-9.21} \\ 
FAJ & \textsf{Portfolio Strategy} & 11.01 & \textsf{Debt Financing} & 0.28 & 2.11 & \grbbb{ 2.50} \\ 
FM & \textsf{Corporate Social Responsibility} & 9.47 & \textsf{Monetary Policy} & 0.36 & 1.51 &  0.52 \\ 
FR & \textsf{Trading/Liquidity} & 7.77 & \textsf{Residential Market} & 0.32 & 1.01 &  0.53 \\ 
FRL & \textsf{Contagion/Spillover} & 7.52 & \textsf{IPO} & 0.56 & 1.08 & \grbbb{-5.45} \\ 
IRFA & \textsf{Contagion/Spillover} & 7.17 & \textsf{Residential Market} & 0.61 & 0.93 & \grbbb{-4.63} \\ 
JB & \textsf{Retail Payment System} & 4.56 & \textsf{Political Relationship/Corruption} & 0.36 & 0.60 & \grbbb{-5.08} \\ 
JBF & \textsf{Bank Liquidity} & 7.74 & \textsf{Political Relationship/Corruption} & 0.69 & 0.68 & \grbbb{-3.09} \\ 
JCF & \textsf{Board/Executive} & 10.17 & \textsf{Interest Rate} & 0.32 & 2.37 & \grbbb{-3.77} \\ 
JD & \textsf{Option Pricing} & 34.11 & \textsf{Corporate Social Responsibility} & 0.23 & 15.10 &  0.75 \\ 
JEF & \textsf{Return Predictability} & 6.96 & \textsf{Political Relationship/Corruption} & 0.31 & 1.67 & \grbbb{-5.96} \\ 
JF & \textsf{Trading/Liquidity} & 6.50 & \textsf{Political Relationship/Corruption} & 0.51 & 0.69 & \grbb{-1.35} \\ 
JFE & \textsf{Trading/Liquidity} & 4.97 & \textsf{Hedging} & 0.63 & 0.62 & \grbbb{-4.16} \\ 
JFI & \textsf{Bank Liquidity} & 13.47 & \textsf{Hedging} & 0.32 & 2.47 & -1.52 \\ 
JFM & \textsf{Volatility} & 19.28 & \textsf{Debt Financing} & 0.12 & 7.82 & \grbb{-3.62} \\ 
JFMKT & \textsf{Trading/Liquidity} & 25.02 & \textsf{Political Relationship/Corruption} & 0.30 & 7.12 & \grbbb{-5.10} \\ 
JFQA & \textsf{Corporate Social Responsibility} & 5.50 & \textsf{Political Relationship/Corruption} & 0.56 & 0.86 & \grbbb{-2.89} \\ 
JFR & \textsf{Trading/Liquidity} & 8.57 & \textsf{Political Relationship/Corruption} & 0.32 & 1.42 & -0.89 \\ 
JFSR & \textsf{Bank Liquidity} & 15.56 & \textsf{Hedging} & 0.33 & 3.15 &  1.38 \\ 
JIFMIM & \textsf{Contagion/Spillover} & 7.91 & \textsf{Payout Policy} & 0.47 & 1.83 & \grbbb{-5.08} \\ 
JIMF & \textsf{Currency} & 20.07 & \textsf{Board/Executive} & 0.22 & 6.82 & \grbbb{-1.77} \\ 
JMCB & \textsf{Monetary Policy} & 30.60 & \textsf{Hedging} & 0.30 & 10.88 & -1.03 \\ 
JPM & \textsf{Portfolio Strategy} & 20.22 & \textsf{Bank Liquidity} & 0.19 & 5.21 &  0.85 \\ 
JREFE & \textsf{Residential Market} & 21.53 & \textsf{Currency} & 0.27 & 6.62 &  0.88 \\ 
JRI & \textsf{Insurance System} & 29.13 & \textsf{Analyst} & 0.27 & 10.98 & -0.85 \\ 
MANF & \textsf{Board/Executive} & 3.31 & \textsf{Bond Market} & 0.41 & 0.56 & -0.82 \\ 
PBFJ & \textsf{Trading/Liquidity} & 5.72 & \textsf{Insurance System} & 0.67 & 0.99 & \grbbb{-3.57} \\ 
QF & \textsf{Option Pricing} & 16.73 & \textsf{Political Relationship/Corruption} & 0.14 & 7.37 & -1.98 \\ 
RF & \textsf{Political Relationship/Corruption} & 6.31 & \textsf{Payout Policy} & 0.64 & 0.62 & \grbbb{-5.32} \\ 
RFS & \textsf{Political Relationship/Corruption} & 8.21 & \textsf{Currency} & 0.65 & 0.93 & -0.66 \\ 
RQFA & \textsf{Residential Market} & 8.07 & \textsf{Political Relationship/Corruption} & 0.49 & 1.17 &  0.32 \\ 
\bottomrule
\end{tabular}}
\end{table} 

\newpage
\begin{table}[H] 
\centering 
\caption{\textbf{Highest- and Lowest-Time-Trend Topics per Journal}\\
The table reports for each journal the two highest and two lowest time-trend topics together with their 
estimated $\beta_k$ in model~\eqref{eq:trendtopic} in parentheses ($\times\!\!$~100). Coefficients are obtained from 
the method of composition described in Section~\ref{sec:testing}.} 
\label{tbl:journal_topic_evol} 
\scalebox{0.55}{ 
\begin{tabular}{lll} 
\toprule
Journal & Highest Time-Trend Topics & Lowest Time-Trend Topics \\ 
\midrule
EFM & \textsf{Corporate Social Responsibility} (1.94), \textsf{Household Credit/Economic Crash} (1.45) & \textsf{Market Anomaly} (-1.24), \textsf{IPO} (-1.12)  \\ 
EJF & \textsf{Household Credit/Economic Crash} (1.93), \textsf{Corporate Social Responsibility} (1.73) & \textsf{Market Anomaly} (-2.28), \textsf{Volatility} (-1.65)  \\ 
FAJ & \textsf{Contagion/Spillover} (3.09), \textsf{Hedging} (2.49) & \textsf{Option Pricing} (-3.22), \textsf{Portfolio Strategy} (-2.17)  \\ 
FM & \textsf{Fund Performance} (1.77), \textsf{Decision-Making Process} (1.63) & \textsf{Market Anomaly} (-1.05), \textsf{IPO} (-0.81)  \\ 
FR & \textsf{Corporate Social Responsibility} (3.47), \textsf{Factor Model} (1.21) & \textsf{Capital Structure} (-1.11), \textsf{Corporate Announcement} (-1.08)  \\ 
FRL & \textsf{Corporate Social Responsibility} (1.85), \textsf{Board/Executive} (1.32) & \textsf{Market Anomaly} (-1.78), \textsf{Currency} (-1.58)  \\ 
IRFA & \textsf{Corporate Social Responsibility} (1.80), \textsf{Contagion/Spillover} (1.52) & \textsf{Interest Rate} (-1.57), \textsf{Market Anomaly} (-1.55)  \\ 
JB & \textsf{Corporate Social Responsibility} (1.36), \textsf{Household Credit/Economic Crash} (1.33) & \textsf{Banking Efficiency} (-1.26), \textsf{Interest Rate} (-1.07)  \\ 
JBF & \textsf{Firm Valuation} (0.89), \textsf{Fund Performance} (0.81) & \textsf{Insurance System} (-2.02), \textsf{Payout Policy} (-1.32)  \\ 
JCF & \textsf{Corporate Social Responsibility} (2.58), \textsf{Household Credit/Economic Crash} (1.74) & \textsf{Ownership} (-2.03), \textsf{IPO} (-1.16)  \\ 
JD & \textsf{Political Relationship/Corruption} (1.67), \textsf{Option Pricing} (1.26) & \textsf{Insurance System} (-0.80), \textsf{Interest Rate} (-0.80)  \\ 
JEF & \textsf{Corporate Social Responsibility} (1.67), \textsf{Household Credit/Economic Crash} (1.60) & \textsf{Volatility} (-2.23), \textsf{Market Anomaly} (-1.84)  \\ 
JF & \textsf{Household Credit/Economic Crash} (2.28), \textsf{Political Relationship/Corruption} (1.38) & \textsf{Market Anomaly} (-1.44), \textsf{Volatility} (-0.94)  \\ 
JFE & \textsf{Corporate Social Responsibility} (2.63), \textsf{Household Credit/Economic Crash} (1.70) & \textsf{Market Anomaly} (-1.61), \textsf{Option Pricing} (-1.32)  \\ 
JFI & \textsf{Political Relationship/Corruption} (5.67), \textsf{Household Credit/Economic Crash} (2.39) & \textsf{Option Pricing} (-2.20), \textsf{Insurance System} (-1.43)  \\ 
JFM & \textsf{Household Credit/Economic Crash} (2.06), \textsf{Contagion/Spillover} (1.36) & \textsf{IPO} (-1.41), \textsf{Corporate Announcement} (-1.11)  \\ 
JFMKT & \textsf{Bank Liquidity} (2.87), \textsf{Household Credit/Economic Crash} (2.67) & \textsf{Insurance System} (-4.71), \textsf{Trading/Liquidity} (-1.41)  \\ 
JFQA & \textsf{Household Credit/Economic Crash} (1.58), \textsf{Contagion/Spillover} (1.56) & \textsf{Trading/Liquidity} (-4.23), \textsf{Real Estate/Bubble} (-1.59)  \\ 
JFR & \textsf{Corporate Social Responsibility} (1.71), \textsf{Household Credit/Economic Crash} (1.51) & \textsf{Market Anomaly} (-2.06), \textsf{Interest Rate} (-1.12)  \\ 
JFSR & \textsf{Residential Market} (3.86), \textsf{Household Credit/Economic Crash} (1.25) & \textsf{Insurance System} (-2.11), \textsf{Market Anomaly} (-1.52)  \\ 
JIFMIM & \textsf{Contagion/Spillover} (1.86), \textsf{Factor Model} (1.46) & \textsf{Volatility} (-1.54), \textsf{Market Anomaly} (-0.96)  \\ 
JIMF & \textsf{Corporate Social Responsibility} (2.08), \textsf{Household Credit/Economic Crash} (1.70) & \textsf{Currency} (-2.67), \textsf{Volatility} (-2.39)  \\ 
JMCB & \textsf{Household Credit/Economic Crash} (2.28), \textsf{Contagion/Spillover} (1.56) & \textsf{Currency} (-1.72), \textsf{Volatility} (-1.66)  \\ 
JPM & \textsf{Household Credit/Economic Crash} (2.97), \textsf{Political Relationship/Corruption} (1.32) & \textsf{Banking Efficiency} (-1.03), \textsf{IPO} (-0.93)  \\ 
JREFE & \textsf{Political Relationship/Corruption} (1.50), \textsf{Hedging} (1.14) & \textsf{Insurance System} (-0.92), \textsf{Market Anomaly} (-0.79)  \\ 
JRI & \textsf{Residential Market} (2.55), \textsf{Household Credit/Economic Crash} (1.26) & \textsf{Insurance System} (-1.09), \textsf{Interest Rate} (-1.03)  \\ 
MANF & \textsf{Political Relationship/Corruption} (5.98), \textsf{Household Credit/Economic Crash} (2.21) & \textsf{Option Pricing} (-1.64), \textsf{Market Anomaly} (-1.45)  \\ 
PBFJ & \textsf{Risk Management} (2.69), \textsf{Household Credit/Economic Crash} (1.29) & \textsf{Insurance System} (-0.97), \textsf{Corporate Announcement} (-0.71)  \\ 
QF & \textsf{Corporate Social Responsibility} (1.01), \textsf{Factor Model} (0.40) & \textsf{Insurance System} (-1.60), \textsf{Currency} (-1.40)  \\ 
RF & \textsf{Corporate Social Responsibility} (2.13), \textsf{Political Relationship/Corruption} (1.54) & \textsf{Market Anomaly} (-2.25), \textsf{Volatility} (-1.51)  \\ 
RFS & \textsf{Household Credit/Economic Crash} (1.18), \textsf{Hedging} (0.97) & \textsf{Option Pricing} (-1.67), \textsf{Market Anomaly} (-1.14)  \\ 
RQFA & \textsf{Residential Market} (3.03), \textsf{Corporate Social Responsibility} (1.80) & \textsf{Market Anomaly} (-1.55), \textsf{Interest Rate} (-1.46)  \\ 
\bottomrule
\end{tabular}}
\end{table} 

\newpage
\begin{table}[H] 
\centering 
\caption{\textbf{Topic Prevalence Sensitivity With Respect to Research Team Characteristics}\\
This table reports the top six positive and negative topic prevalence
sensitivity coefficients in model~\eqref{eq:topuniv} with respect to the number of authors (Panel~A, coefficient $c_k$), the presence of at least one female (Panel~B, coefficient $d_k$), 
and the presence of at least one top-tier institution scholar (Panel~C, coefficient $e_k$). Coefficients and p-values are obtained from 
the method of composition described in Section~\ref{sec:testing}. All coefficients are multiplied by 100 for readability purposes.}
\label{tbl:team} 
\scalebox{0.95}{ 
\begin{tabular}{llll} 
\toprule
\multicolumn{4}{l}{Panel A: Number of Authors}\\
\multicolumn{2}{c}{Positive Coefficients} & \multicolumn{2}{c}{Negative Coefficients} \\ 
\cmidrule(lr){1-2}\cmidrule(lr){3-4}
Topic & $c_k$ & Topic & $c_k$ \\
\midrule
\textsf{Corporate Announcement} &  \grbbb{3.37} &   \textsf{Monetary Policy} &  \grbbb{-8.02} \\
\textsf{Market Anomaly} &  \grbbb{3.00} &   \textsf{Retail Payment System} &  \grbbb{-4.50} \\
\textsf{Disclosure} &  \grbbb{2.82} &   \textsf{Interest Rate} &  \grbbb{-3.69} \\
\textsf{Emerging Country} &  \grbbb{2.79} &   \textsf{Capital Structure} &  \grbbb{-3.06} \\
\textsf{Contagion/Spillover} &  \grbbb{2.61} &   \textsf{Currency} &  \grbbb{-2.95} \\
\textsf{Factor Model} &  \grbbb{2.17} &   \textsf{Corporate Investment} &  \grbbb{-2.75} \\
\midrule
\multicolumn{4}{l}{Panel B: At Least One Female Author}\\
\multicolumn{2}{c}{Positive Coefficients} & \multicolumn{2}{c}{Negative Coefficients} \\ 
\cmidrule(lr){1-2}\cmidrule(lr){3-4}
Topic & $d_k$ & Topic & $d_k$ \\
\midrule
\textsf{Corporate Social Responsibility} &  \grbbb{10.84} &  \textsf{Return Predictability} &  \grbbb{-9.60} \\
\textsf{Board/Executive} &  \grbbb{8.46} &  \textsf{Portfolio Strategy} &  \grbbb{-9.38} \\
\textsf{Ownership} &  \grbbb{6.29} &  \textsf{Interest Rate} &  \grbbb{-8.30} \\
\textsf{Corporate Announcement} &  \grbbb{5.42} &   \textsf{Volatility} &  \grbbb{-7.05} \\
\textsf{Emerging Country} &  \grbbb{5.39} &  \textsf{Market Anomaly} &  \grbbb{-6.00} \\
\textsf{Bank Liquidity} &  \grbbb{5.07} &  \textsf{Monetary Policy} &  \grbbb{-5.63} \\
\midrule
\multicolumn{4}{l}{Panel C: At Least One Top-Tier Institution Author}\\
\multicolumn{2}{c}{Positive Coefficients} & \multicolumn{2}{c}{Negative Coefficients} \\ 
\cmidrule(lr){1-2}\cmidrule(lr){3-4}
Topic & $e_k$ & Topic & $e_k$ \\
\midrule
\textsf{Decision-Making Process} &  \grbbb{10.64} &   \textsf{Contagion/Spillover} &  \grbbb{-6.49} \\
\textsf{Corporate Investment} &  \grbbb{10.59} &   \textsf{Emerging Country} &  \grbbb{-5.97} \\
\textsf{Debt Financing} &  \grbbb{8.70} & \textsf{Volatility} &  \grbbb{-3.69} \\
\textsf{Capital Structure} &  \grbbb{8.65} &   \textsf{Market Anomaly} &  \grbbb{-3.63} \\
\textsf{Political Relationship/Corruption} &  \grbbb{8.36} &   \textsf{Disclosure} &  \grbb{-3.44} \\
\textsf{Fund Performance} &  \grbbb{7.34} &   \textsf{Currency} &  \grbb{-3.38} \\
\bottomrule
\end{tabular}}
\end{table} 

\newpage

\newpage

\newpage

\newpage

\newpage

\newpage

\newpage

\newpage

\newpage
\begin{titlepage}
\begin{center}
\vspace*{1cm}
\huge{
Thirty Years of Academic Finance\\[.5cm]
Online Appendix}\\   
\vspace{1.5cm}
\Large{
David Ardia$^{1}$,
Keven Bluteau$^{2}$,
Mohammad-Abbas Meghani$^{1}$}\\[.5cm]
\normalsize{
$^{1}$GERAD \& Department of Decision Sciences, HEC Montr\'eal, Canada\\
$^{2}$Department of Finance, University of Sherbrooke, Canada}
\vspace{1.5cm}\\
\Large{\today}
\vfill
\end{center}
\end{titlepage}

\pagenumbering{arabic} 
\linespread{1.2}
\onehalfspacing

\setlength\parindent{0pt}

\setcounter{section}{0}
\renewcommand*{\thesection}{\Alph{section}} 

\setcounter{equation}{0}
\renewcommand*{\theequation}{\thesection.\arabic{equation}} 

\setcounter{table}{0}
\renewcommand*{\thetable}{\thesection.\arabic{table}} 

\setcounter{figure}{0}
\renewcommand*{\thefigure}{\thesection.\arabic{figure}} 

\setcounter{page}{1}
\renewcommand*{\thepage}{\arabic{page}} 

\newpage
\section{Topic labeling}
\label{app:topiclabel}

\setstretch{0.8}
\footnotesize
\begin{longtable}{p{4cm}p{10cm}}
\label{tab:topiclabel}\\
\caption{\textbf{Topic Labeling}\\
This table lists the 50 topics with the ten-largest prevalent keywords. The topics are sorted in alphabetical order.}\\
\toprule
Topic & Keywords  \\
\endfirsthead
\multicolumn{2}{c}%
{\tablename\ \thetable\ -- \textit{Continued from previous page}} \\
\toprule
Topic &  Keywords \\
\midrule
\endhead
\bottomrule\multicolumn{2}{r}{\textit{Continued on next page}} \\
\endfoot
\bottomrule
\endlastfoot
\midrule
\textsf{Analyst} & analyst, bia, institutional\_investor, dispersion, forecast, recommendation, side, revision, institutional\_ownership, earning\_forecast \\ 
\textsf{Bank Liquidity} & bank, banks, banking, capital, liquidity, systemic\_risk, funding, paper, deposit, effect \\ 
\textsf{Banking Efficiency} & efficiency, technology, production, productivity, cost, stage, scale, revenue, inefficiency, performance \\ 
\textsf{Board/Executive} & board, ceo, compensation, incentive, ceos, performance, director, firm\_performance, family, firm \\ 
\textsf{Bond Market} & bond, spread, yield, cds, bond\_market, corporate\_bond, call, credit\_spread, credit\_default\_swap, government\_bond \\ 
\textsf{Capital Structure} & leverage, asset, capital\_structure, bankruptcy, firm, off, adjustment, theory, financial\_distress, debt \\ 
\textsf{Concentration/Competition} & index, measure, competition, sector, concentration, entry, market\_power, deregulation, inclusion, result \\ 
\textsf{Contagion/Spillover} & stock\_market, crisis, correlation, period, financial\_crisis, spillover, contagion, financial\_market, oil, time \\
\end{longtable}%

\newpage
\section{Vocabulary}
\label{sec:vocabulary}

\tiny

\setstretch{0.7}
\begin{longtable}{p{4cm}p{1cm}p{1cm}p{4cm}p{1cm}p{1cm}}
\caption{\textbf{Vocabulary}\\}\\
\toprule
Token &  Freq & \% of doc &	Token &  Freq & \%  of doc \\
\endfirsthead
\multicolumn{3}{c}%
{\tablename\ \thetable\ -- \textit{Continued from previous page}} \\
\toprule
Token  &  Freq &  \%  of doc &	Token   &  Freq  \%  of doc &\\
\midrule
\endhead
\bottomrule\multicolumn{3}{r}{\textit{Continued on next page}} \\
\endfoot
\bottomrule
\endlastfoot
\midrule
abilitie &  71 &  0.12 &  justification &  78 &  0.15\\
ability &  2,045 &  3.63 &  kalman &  126 &  0.19\\
abnormal\_return &  1,344 &  2.07 &  keiretsu &  103 &  0.1\\
abnormal\_stock\_return &  161 &  0.32 &  kernel &  158 &  0.2\\
absence &  446 &  0.92 &  key &  54 &  0.11\\
abuse &  55 &  0.09 &  keynesian &  139 &  0.26\\
academic &  751 &  1.63 &  kind &  189 &  0.4\\
academics &  128 &  0.26 &  kingdom &  289 &  0.52\\
\end{longtable}%
\end{document}